\documentclass[twocolumn,showpacs,preprintnumbers,amsmath,amssymb,APSl,nofootinbib,superscriptaddress]{revtex4-1}
\usepackage{bm}
\usepackage{mathrsfs}
\usepackage{xcolor,color,graphicx,graphics}
\usepackage[all]{xy}
\usepackage{latexsym,amssymb,amsmath,amsfonts} %\usepackage{amstex}
\usepackage[english]{babel} %, spanish, portuguese
\usepackage[OT1]{fontenc}
\usepackage[latin1]{inputenc}
\usepackage{makeidx}
\usepackage{hyperref}
\definecolor{red}{rgb}{1,0,0}

\def\+{^\dagger}

\def\<{\leftarrow}
\def\>{\rightarrow}

\def\({\left(}
\def\){\right)}

\newcommand{\bi}{\begin{itemize}} 				\newcommand{\ei}{\end{itemize}}
\newcommand{\benu}{\begin{enumerate}} 		\newcommand{\enu}{\end{enumerate}}
\newcommand{\bd}{\begin{dinglist}{0}}     \newcommand{\ed}{\end{dinglist}}
\newcommand{\bfig}{\begin{figure}[htbp]}  \newcommand{\efig}{\end{figure}}
        			
\newcommand{\bc}{\begin{center}} 				  \newcommand{\ec}{\end{center}}
\newcommand{\be}{\begin{equation}} 				\newcommand{\ee}{\end{equation}}
\newcommand{\bsub}{\begin{subequations}}  \newcommand{\esub}{\end{subequations}}
\newcommand{\ben}{\begin{eqnarray}} 			\newcommand{\een}{\end{eqnarray}}
\newcommand{\ba}[1]{\begin{array}{#1}} 		\newcommand{\ea}{\end{array}}
\newcommand{\bea}{\begin{equation}\begin{array}{rcl}}
\newcommand{\eea}{\end{array}\end{equation}}

\begin{document}

\title{Einstein-Cartan-Dirac gravity with $U(1)$ symmetry breaking}

\author{Francisco Cabral}
\email{cosmocabral@gmail.com}
\affiliation{Instituto de Astrof\'{\i}sica e Ci\^{e}ncias do Espa\c{c}o, Faculdade de
Ci\^encias da Universidade de Lisboa, Edif\'{\i}cio C8, Campo Grande,
P-1749-016 Lisbon, Portugal}
\author{Francisco S. N. Lobo} \email{fslobo@fc.ul.pt}
\affiliation{Instituto de Astrof\'{\i}sica e Ci\^{e}ncias do Espa\c{c}o, Faculdade de
Ci\^encias da Universidade de Lisboa, Edif\'{\i}cio C8, Campo Grande,
P-1749-016 Lisbon, Portugal}
\author{Diego Rubiera-Garcia} \email{drubiera@ucm.es}
\affiliation{Departamento de F\'isica Te\'orica and IPARCOS, Universidad Complutense de Madrid, E-28040
Madrid, Spain}
\affiliation{Instituto de Astrof\'{\i}sica e Ci\^{e}ncias do Espa\c{c}o, Faculdade de
Ci\^encias da Universidade de Lisboa, Edif\'{\i}cio C8, Campo Grande,
P-1749-016 Lisbon, Portugal}

\date{\today}

%%%%%%%%%%%%%%%%%%%%%%%%%%%%%%%%%%%%%%%%%%%%%%%%%%%%%%%%%%%%%%%%%%%%%%%%
\begin{abstract}
Einstein-Cartan theory is an extension of the standard formulation of General Relativity where torsion (the antisymmetric part of the affine connection) is non-vanishing. Just as the space-time metric is sourced by the stress-energy tensor of the matter fields,  torsion is sourced via the spin density tensor, whose physical effects become relevant at very high spin densities. In this work we introduce an extension of the Einstein-Cartan-Dirac theory  with an electromagnetic (Maxwell) contribution minimally coupled to torsion. This contribution breaks the $U(1)$ gauge symmetry, which is suggested by the possibility of  a torsion-induced phase transition in the early Universe, yielding new physics in extreme (spin) density regimes.  We obtain the generalized gravitational, electromagnetic and fermionic field equations for this theory, estimate the strength of the corrections, and discuss the corresponding phenomenology. In particular, we briefly address some astrophysical considerations regarding the relevance of the effects which might take place inside ultra-dense neutron stars with strong magnetic fields (magnetars).
% The cosmological applications of this theory in the very early Universe, including bounces, generation of an effective cosmological constant and matter-antimatter asymmetry in baryogenesis are rigorously explored in a separate work. 
\end{abstract}
%%%%%%%%%%%%%%%%%%%%%%%%%%%%%%%%%%%%%%%%%%%%%%%%%%%%%%%%%%%%%%%%%%%%%%%%

%%%%%%%%%%%%%%%%%%%%%%%%%%%%%%%%%%%%%%%%%%%%%%%%%%%%%%%%%%%%%%%%%%%%%%%%
%\keywords{exterior calculus of forms, Riamann-Cartan space-time, torsion, electrodynamics, gravity waves}
%%%%%%%%%%%%%%%%%%%%%%%%%%%%%%%%%%%%%%%%%%%%%%%%%%%%%%%%%%%%%%%%%%%%%%%%

\maketitle

%%%%%%%%%%%%%%%%%%%%%%%%%%%%%%%%%%%%%%%%%%%%%%%%%%%%%%%%%%%%%%%%%%%%%%%%
\section{Introduction}
%%%%%%%%%%%%%%%%%%%%%%%%%%%%%%%%%%%%%%%%%%%%%%%%%%%%%%%%%%%%%%%%%%%%%%%%

We have recently witnessed the birth of gravitational wave astronomy, where the LIGO/VIRGO Collaboration reported compelling evidence on the detection of gravitational waves, which is compatible with a scenario of binary black hole mergers predicted by  General Relativity (GR) \cite{Abbott:2016blz,Abbott:2017nn}. This finding has sparked the interest in probing the strong-field regime of GR via gravitational wave observations of compact objects \cite{TheLIGOScientific:2016src}. At the writing of this paper around a half-dozen events have been detected including a neutron star binary merger  \cite{Abbott:2017oio}, with its electromagnetic counterpart compatible with a short gamma ray burst \cite{TheLIGOScientific:2017qsa}. In the aftermath of these observations, researchers have quickly gone to discuss how well their favourite gravitational models extending GR have fared against them \cite{Lombriser:2015sxa, Lombriser:2016yzn, Baker:2017hug,Sakstein:2017xjx,Creminelli:2017sry,Ezquiaga:2017ekz,EzZuma, Copeland:2018yuh}. But given the undeniable  experimental success of GR \cite{Will:2014kxa} why would one consider going beyond it? Modified theories of gravity are indeed motivated by a variety of reasons. On the observational side they are introduced as alternatives to dark matter/dark energy scenarios. On the theoretical side, the need of an ultraviolet completion of GR and the unavoidable existence of space-time singularities deep inside black holes and in the early Universe have been troubling researchers for decades. For some reviews on the motivations of these theories and their phenomenology, see e.g. \cite{DeFelice:2010aj,Olmoreview, CLreview,NOOreview,Berti:2015itd,BeltranJimenez:2017doy,Harko:2014gwa,Capozziello:2015lza,BookHarkoLobo}.

The traditional approach to formulate GR is to consider a symmetric rank-two metric tensor $g_{\mu\nu}$ endowed with a pseudo-Riemannian manifold. Parallel transport is mediated by an affine connection $\Gamma^{\alpha}_{\mu\nu}$, which is assumed to be symmetric, $\Gamma^{\alpha}_{\mu\nu}=\Gamma^{\alpha}_{\nu\mu}$, and metric-compatible, $\nabla_{\alpha}^{\Gamma} g_{\mu\nu}=0$, with a minimal coupling between matter fields and gravity ensuring that the equivalence principle (i.e., the universality of free-fall) is fulfilled. However, it is known that a manifold equipped with zero curvature and non-metricity but non-zero torsion (also known as
Weitzenb\"{o}ck space-time) allows to build a theory of gravity that is fully equivalent to GR when the linear action in the torsion scalar is chosen. This is known as the teleparallel equivalent of General Relativity 
\cite{Hayashi:1979qx,JGPereira,Maluf:2013gaa,Bamba:2013jqa}.
There is yet another formulation fully equivalent to GR, based on zero curvature and torsion but non-zero non-metricity, called symmetric teleparallel gravity, whose properties have begun to be unravelled very recently \cite{BeltranJimenez:2017tkd,BeltranJimenez:2018vdo,Jarv:2018bgs,Iosifidis:2018zjj,Harko:2018gxr,Lazkoz:2019sjl}.

Given these equivalences in gravitational models under different space-time paradigms, it is natural to ask if there is any guiding principle. Specifically, what space-time geometry and fundamental degrees of freedom can represent gravity? Indeed, the study of non-Riemann geometries is connected to the topic of gauge symmetries. In the gauge formulation of gravitation, the gauge principle is applied to space-time symmetries leading to non-Riemann geometrical structures such as torsion and non-metricity (depending on the local symmetry group). This principle can provide guidance for the study of the role of non-Riemann geometries in gravitation and in unified field theories, in connection to symmetries in physics and symmetry breaking phase transitions (for detailed reviews on gauge theories of gravity and its applications see \cite{Blagojevic:2013xpa,Blagojevic,Obukhov} and references therein). The history of the gauge principle is actually very rich, dating back one hundred years to the original proposal by Weyl in his unified field theory of gravity (and electromagnetism). For this purpose, Weyl introduced a space-time (Weyl geometry) with curvature and a non-vanishing trace-vector part of the non-metricity \cite{Unified field}. Later, Kibble gauged the Poincar\'{e} global symmetries of Minkowski space-time to arrive at the Einstein-Cartan-Sciama-Kibble theory of gravity (EC for short), within a space-time with curvature and torsion (Riemann-Cartan geometry, or RC). Poincar\'{e} gauge theories of gravity (PGTG) have been investigated with special interest on Lagrangians quadratic in the curvature and torsion invariants, and in applications regarding cosmology, gravitational waves, and spherical solutions, see e.g. \cite{Obukhov:2018bmf,Obukhov:2017pxa,Blagojevic:2018dpz,Blagojevic:2015zma,Cembranos:2019mcb}.

In this work we shall focus on EC theory \cite{Hehl:2007bn} (the simplest example of PGTGs), which allows to consistently incorporate the intrinsic angular momentum (spin) of fermionic matter. Thus, in EC theory, in addition to the stress-energy matter sources, the spin energy density is also a source of the gravitational field. Torsion in EC theory becomes important in scenarios where high-spin densities are present. Although Cartan introduced the theory almost one century ago, it continues to trigger interest due to its non-singular solutions (in black holes and cosmology), by the bridge it establishes between fermionic spinors and gravity, and by its elegance and simplicity, as it possesses no free parameters besides Newton's constant. Within its many applications we underline bouncing cosmologies \cite{Poplawski:2011jz,Unger:2018oqo,Kranas:2018jdc}, inflation, cosmological constant and dark energy \cite{Poplawski:2010kb,Ivanov:2016xjm,Razina:2010bj}, perturbations and cosmic microwave background radiation \cite{Palle:2014goa, Poplawski:2012qy}, phase and signature transitions \cite{Xue:2008qs,Vakili:2013fra}, or compact objects \cite{Bronnikov:2016xvj,Cembranos:2017pcs}.
For EC gravity coupled to Dirac fields, one also finds applications in particle physics, see e.g. \cite{Khanapurkar:2018jvx,Khanapurkar:2018gyo,Lucat:2015rla,Poplawski:2011wj,Poplawski:2011xf,Poplawski:2010jv,Poplawski:2009su}.

The need to go beyond EC theory was recognized long ago, mainly due to the fact that the theory is still non-renormalizable \cite{Shapiro:2014kma}, but also because quadratic Lagrangians present a natural and theoretically preferable extension \cite{Baekler:2011jt,Fabbri:2012qr,Baekler:2010fr} (see also \cite{Obukhov:2012je,Xue:2011qy}). A path recently explored to extend this theory is the analysis of new (non-minimal) couplings between  torsion and the matter fields \cite{Harko:2014sja,Harko:2014aja,Carloni:2015lsa,Gonzalez-Espinoza:2018gyl,Jawad:2016zwj}. In this sense, the coupling between torsion and electromagnetism has been carefully analysed in the literature with the result that, in general, this can be achieved by either changing the field equations with minimal/non-minimal couplings, or via the constitutive relations between the field strengths $\mathit{F}=(E,B)$ and the excitations $\mathit{H}=(D,H)$ (see \cite{FHehl:2001,Rubilar:2003}). Though it is usually assumed that torsion does not minimally couple to the electromagnetic field, since this breaks the $U(1)$ gauge invariance (for details see e.g. \cite{Shapiro:2014kma,FHehl:2001}), the physics of phase transitions in condensed matter systems, superconductivity and early universe is permeated by processes that lead to spontaneous symmetry breaking, and in high density environments torsion can provide a physical mechanism to induce such a symmetry breaking. Since in EC theory torsion vanishes outside the matter sources and is negligible at low densities, the Maxwell equations remain valid for all phenomena that we can presently probe directly.

The main aim of this work is to study the (minimal) coupling of torsion with fermions and electromagnetism (massless spin  one bosonic fields). Through the Cartan equations relating the space-time torsion to the matter fields, this coupling induces non-minimal and self-interactions of the matter fields, and provides a physical mechanism to generate a $U(1)$ symmetry breaking for high densities and fields. In the broken phase, torsion provides an effective mass for the photon, with the electromagnetic potential obeying an extended Proca-like equation. 

%Comparing with the analysis of Ref. \cite{Poplawski:2011cr},

In our approach we consider first the regime in which torsion is only sourced by fermions, and extend it later to the general case where both fermionic and bosonic fields contribute to torsion via the corresponding spin energy densities. The first case is a simplifying ansatz, where the electromagnetic fields are influenced by the (background) space-time torsion but do not backreact on it. This case serves to illustrate some of the effects in the new dynamics due to the minimal coupling of bosons and torsion. The second case encodes the full dynamics with the bosonic spin contribution to torsion, which induces new non-linearities. Therefore, in this paper we address the most relevant features of EC-Dirac-Maxwell model with $U(1)$ symmetry breaking as well as some of its physical implications.
%The cosmological applications of this model are explored in a separate work.

This paper is organized as follows. In Sec. \ref{secII} we review EC gravity, focusing on the case where fermions are represented by a Dirac field. Sec. \ref{secIII} contains the core results of this work, where we extend EC theory by introducing the electromagnetic field minimally coupled to torsion, and find the corresponding dynamics for gravitational, electromagnetic and fermionic sectors in two cases: (i) fermionic background torsion and (ii) the full case, including the bosonic backreaction to torsion via its spin tensor. We conclude in Sec. \ref{secIV} with a broad discussion of the phenomenological implications of these results, including some future perspectives.

\section{Einstein-Cartan theory with fermions}\label{secII}

\subsection{Einstein-Cartan gravity}

In general, any affine connection can be decomposed into three independent pieces
\begin{equation} \label{eq:affcons}
\Gamma^{\lambda}{}_{\mu\nu} =
	\tilde{\Gamma}^{\lambda}{}_{\mu\nu} +
	K^{\lambda}{}_{\mu\nu}+
	L^{\lambda}{}_{\mu\nu}\,,
\end{equation}
where $\tilde{\Gamma}^{\lambda}{}_{\mu\nu}$ is the Levi-Civita connection associated to the Riemannian curvature ${\tilde{R}^\alpha}{}_{\beta\mu\nu}=2\partial_{[\mu}{\tilde{\Gamma}^\alpha}{}_{\nu]\beta} +2 {\tilde{\Gamma}^\alpha}{}_{[\mu \vert \lambda \vert}{\tilde{\Gamma}^\lambda}{}_{\nu ] \beta}$, the second term is associated to torsion $T^{\lambda}{}_{\alpha\beta}\equiv\Gamma^{\lambda}{}_{[\alpha\beta]}$ and is denoted contortion\footnote{By construction, contortion is antisymmetric on its first two indices, $K_{\alpha \beta \gamma}=-K_{\beta \alpha \gamma}$.}
\begin{equation} \label{eq:contor}
K^{\lambda}{}_{\mu\nu} \equiv  T^{\lambda}{}_{\mu\nu}-2T_{(\mu}{}^{\lambda}{}_{\nu)},
\end{equation}
while the third term is associated to non-metricity $Q_{\rho \mu \nu} \equiv \nabla_{\rho} g_{\mu\nu}$ and is called disformation,
\begin{equation}
L^{\lambda}{}_{\mu\nu} \equiv \frac{1}{2} g^{\lambda \beta} \left( -Q_{\mu \beta\nu}-Q_{\nu \beta\mu}+Q_{\beta \mu \nu} \right) \ .
\end{equation}
Throughout this paper all objects with tilde will refer to expressions computed in a Riemannian space-time.

From the general decomposition (\ref{eq:affcons}), keeping curvature and torsion but setting non-metricity to zero we obtain a Riemann-Cartan (RC) space-time geometry. In this case the curvature scalars in the Cartan connection and the Levi-Civita connection are related as
\begin{equation}
R=\tilde{R}-2\tilde{\nabla}^{\lambda}K^{\alpha}{}_{\lambda\alpha} +g^{\beta\nu}\left(K^{\alpha}{}_{\lambda\alpha}K^{\lambda}{}_{\beta\nu}-K^{\alpha}{}_{\lambda\nu}K^{\lambda}{}_{\beta\alpha}\right).
\end{equation}
Now, selecting the linear Lagrangian in the curvature scalar yields the action
\begin{equation} \label{eq:actionEC}
S_{\rm EC}=\dfrac{1}{2\kappa^2} \int d^4x \sqrt{-g}R(\Gamma) + \int d^4x \sqrt{-g}\,\mathcal{L}_{m}\,,
\end{equation}
with the following definitions and conventions: $\kappa^2=8\pi G$ is Newton's constant, $g$ is the determinant of the space-time metric $g_{\mu\nu}$, the curvature scalar $R=g_{\mu\nu}R^{\mu\nu}$ is constructed out of the Ricci tensor $R_{\mu\nu}(\Gamma) \equiv {R^\alpha}_{\mu\alpha\nu}(\Gamma)$, and the matter Lagrangian, $\mathcal{L}_m=\mathcal{L}_{m}(g_{\mu\nu},\Gamma,\psi_m)$, depends on the metric and the matter fields, collectively denoted by $\psi_m$, and also on the contortion via the covariant derivatives. The above action resembles GR, but the fact that the connection has now an antisymmetric part yields new contributions to the standard Einstein equations and to the dynamical equations for the matter fields, as we shall see below.

To obtain the field equations for EC theory we start by varying the  action (\ref{eq:actionEC}) with respect to the contortion tensor $K^{\alpha}_{\;\beta\gamma}$, which yields the so-called Cartan equations
\begin{equation}
T^{\alpha}_{\;\beta\gamma}+T_{\gamma}\delta^{\alpha}_{\beta}-T_{\beta}\delta^{\alpha}_{\gamma}=\kappa^2 s^{\alpha}_{\;\beta\gamma},\label{cartaneqs}
\end{equation}
or
\begin{equation}
T^{\alpha}_{\;\beta\gamma}=\kappa^2 \left( s^{\alpha}_{\;\beta\gamma}+\delta^{\alpha}_{[\beta}s_{\gamma]}\right),\nonumber \\
\end{equation}
where
\begin{equation}
\label{spintensor}
s^{\gamma\alpha\beta}\equiv\dfrac{\delta \mathcal{L}_{m}}{\delta K_{\alpha\beta\gamma}}\,,
\end{equation}
is the spin density tensor with dimensions of energy/area, $T_{\beta}\equiv T^{\gamma}_{\;\beta\gamma}$ and $s_{\beta}\equiv s^{\gamma}_{\;\beta\gamma}$ are the torsion and spin (trace) vectors, respectively\footnote{It is useful to express the torsion tensor in terms of its irreducible components $T^{\lambda}_{\;\mu\nu}=\bar{T}^{\lambda}_{\;\mu\nu}+\dfrac{2}{3}\delta^{\lambda}_{[\nu}T_{\mu]}+g^{\lambda\sigma}\epsilon_{\mu\nu\sigma\rho} \breve{T}^{\rho}$, where the traceless tensor obeys $\bar{T}^{\lambda}_{\;\mu\lambda}=0,\quad \epsilon^{\lambda\mu\nu\rho}\bar{T}_{\mu\nu\rho}=0$, $T_{\mu}$ is the trace vector and $\breve{T}^{\lambda}\equiv \dfrac{1}{6}\epsilon^{\lambda\alpha\beta\gamma}T_{\alpha\beta\gamma}$ is the  pseudo-trace (axial) vector.}. Cartan's equations (\ref{cartaneqs}) imply that, analogously to curvature being sourced by the stress-energy of the matter sources, torsion is sourced by its spin density. These  equations are linear  and algebraic, which implies that outside regions with spin densities (and, in particular, in vacuum) they vanish identically.

The Lagrangian in (\ref{eq:actionEC}) can be expressed in the following way
\begin{equation} \label{eq:LagrangEC}
\mathcal{L}_{\rm EC}=\dfrac{1}{2\kappa^2}\tilde{R} +\mathcal{L}^{eff}_{m}\,,
\end{equation}
with
\begin{equation}
\mathcal{L}^{\rm eff}_{m}=\mathcal{L}_{m}-\dfrac{1}{2\kappa^2}
\Big(K^{\alpha\lambda}_{\;\;\;\alpha}K_{\gamma\lambda}^{\;\;\;\gamma}+K^{\alpha\lambda\beta}K_{\lambda\beta\alpha}\Big), 
\end{equation}
or, explicitly
\begin{equation}
\mathcal{L}^{\rm eff}_{m}=\mathcal{L}_{m}-\dfrac{\kappa^{2}}{2}\left[s^{\lambda}s_{\lambda}+s^{\mu\nu\lambda}\left(s_{\nu\lambda\mu}+s_{\lambda\mu\nu}+s_{\mu\lambda\nu}\right)\right]\nonumber \\, 
\end{equation}
where we used Cartan's equations in the second term and we have neglected the surface term which does not contribute to the field equations. Therefore, variation of the action (\ref{eq:actionEC}) with respect to the space-time metric $g_{\mu\nu}$ yields the generalized Einstein equations, which can be suitably written as
\begin{equation}
\tilde{G}_{\mu\nu}=\kappa^2(T_{\mu\nu}+U_{\mu\nu}),\label{ECeqs}
\end{equation}
where $\tilde{G}_{\mu\nu}$ is the Einstein tensor computed with the Levi-Civita connection in Eq. (\ref{eq:affcons}). On the right-hand side, we have the effective stress-energy tensor 
\begin{equation}
T^{\rm eff}_{\mu\nu}=T_{\mu\nu}+U_{\mu\nu}=\frac{2}{\sqrt{-g}} \frac{\delta (\sqrt{-g}\mathcal{L}^{\rm eff}_{m})}{\delta g^{\mu\nu}}\,,
\end{equation}
which  we split  in the stress-energy tensor of the matter fields $T_{\mu\nu}=\frac{2}{\sqrt{-g}} \frac{\delta (\sqrt{-g}\mathcal{L}_m)}{\delta g^{\mu\nu}}$, plus another tensor, $U_{\mu\nu}=\frac{2}{\sqrt{-g}} \frac{\delta (\sqrt{-g}C)}{\delta g^{\mu\nu}}$, with $C\equiv-\dfrac{1}{2\kappa^2}
\Big(K^{\lambda}K_{\lambda}+K^{\alpha\lambda\beta}K_{\lambda\beta\alpha}\Big)$, which contains the corrections quadratic in torsion $U\sim\kappa^{-2}T^{2}$ or in the spin variables $U\sim \kappa^2 s^2$, via Cartan's equations (\ref{cartaneqs}). It is important to point out that, in general, torsion also contributes to the stress-energy tensor $T_{\mu\nu}$, since the covariant derivatives present in the kinetic part of $\mathcal{L}_m$ introduce new terms depending on torsion via minimal couplings (non-minimal couplings can also be present).
Since $U\sim \kappa^2 s^2$, Eq. (\ref{ECeqs}) defines a typical density $\rho_{C} \sim 10^{54}$g/cm$^3$, known as Cartan's density. This is much higher than the nuclear saturation density,  $\rho_s \sim 10^{14}$g/cm$^3$,
though much lower than Planck's density $\rho_P \sim 10^{93}$ g/cm$^3$, where quantum corrections to classical gravitation are expected to arise. Therefore, in principle, EC theory can only introduce significant physical effects in environments of very large spin densities, which might arise in the early universe or in the innermost regions of black holes. 

Let us also note that, though the contribution in $U_{\mu\nu}$ from the spin density to the space-time metric (or curvature) is only significant at or above the Cartan density $\rho_{C}$, one can show that torsion-induced effects are predicted at smaller densities in the dynamics of fermions (via the so-called Hehl-Datta term, as discussed below). Therefore, the physics of EC-Dirac systems, as well as in those where an additional electromagnetic contribution is considered (as we shall see in this work) should also be studied inside ultra-compact objects such as neutron stars, particularly in magnetars, and even in quark stars. It should also be mentioned that  Cartan's energy scale is not always the correct scale to guide the physical reasoning. This is clear for elementary standard model fermions since, as shown in \cite{Diether:2017oax}, there are important (compensating) physical mechanisms at very small distances (Planck's scale, not Cartan's distance scale which is derived from Cartan's density) which allow the regularization of the self-energy densities around the electroweak scales (in the case of leptons), corresponding to the observed masses ($m_{e}, m_{\mu}, m_{\tau},...$).

Another aspect of the EC equations (\ref{cartaneqs})--(\ref{ECeqs}) worth emphasizing is that, in the absence of spin density, they boil down to the GR ones. Therefore, EC theory in vacuum does not propagate additional degrees of freedom beyond the two tensor polarizations of the gravitational field  that propagate at the speed of light, thus being in agreement with the recent findings resulting from the LIGO-VIRGO Collaboration on the equality of the speed of light and of gravitational waves \cite{TheLIGOScientific:2017qsa}. Nonetheless, physical mechanisms for generating gravitational waves in the very early universe can also have contributions from spin energy tensor fluctuations around the Cartan density, for example via its effective non-vanishing and time-varying quadrupole moment, or due to phase transitions (including symmetry breaking mechanisms induced by torsion effects)

\subsection{Einstein-Cartan-Dirac theory}

Let us consider, as the matter sector in the action (\ref{eq:actionEC}), a free Dirac fermionic field with mass $m$ minimally coupled to torsion. The corresponding Lagrangian density can be expressed as \cite{Shapiro:2001rz}
\begin{equation}\label{eq:mattfer}
\mathcal{L}_{\rm Dirac}=\dfrac{i\hbar}{2}\left(\bar{\psi}\gamma^{\mu}D_{\mu}\psi-(D_{\mu}\bar{\psi})\gamma^{\mu}\psi\right)-m\bar{\psi}\psi,
\end{equation}
for spinors $\psi$ and their adjoints $\bar{\psi}=\psi^{+}\gamma^0$, and where the covariant derivatives are defined as
\begin{eqnarray}
D_{\mu}\psi&=&\tilde{D}_{\mu}\psi+\dfrac{1}{4}K_{\alpha\beta\mu}\gamma^{\alpha}\gamma^{\beta}\psi , \\
D_{\mu}\bar{\psi}&=&\tilde{D}_{\mu}\bar{\psi}-\dfrac{1}{4}K_{\alpha\beta\mu}\bar{\psi}\gamma^{\alpha}\gamma^{\beta},
\end{eqnarray}
where $D_{\mu}$ and $\tilde{D}_{\mu}$  are the (Fock-Ivanenko) covariant derivatives built with the Cartan connection and the Levi-Civita connection, respectively, and $\gamma^{\mu}$ are the induced Dirac-Pauli matrices\footnote{The usual constant Pauli-Dirac matrices $\gamma^{c}$, which obey $\left\lbrace \gamma^{a},\gamma^{b}\right\rbrace =2\eta^{ab}I$, are related to the $\gamma^{\mu}$ matrices via $\gamma^{\mu}e^{a}_{\;\;\mu}=\gamma^{a}$, where $e^{a}_{\;\;\mu}$ are the tetrads satisfying $g_{\mu\nu}=\eta_{ab}e^{a}_{\;\;\mu}e^{b}_{\;\;\nu}$ and $e^{a}_{\;\;\mu}e^{\;\;\mu}_{b}=\delta^{a}_{b}$, $e^{c}_{\;\;\nu}e^{\;\;\mu}_{c}=\delta^{\mu}_{\nu}$ and $\eta_{ab}$ is the Minkowski metric.} obeying $\left\lbrace \gamma^{\mu},\gamma^{\nu}\right\rbrace =2g^{\mu\nu}I$, where $I$ is the $4\times4$ unit matrix and $g_{\mu\nu}$ is the space-time metric\footnote{The Fock-Ivanenko covariant derivatives of spinors, in Riemann geometry, are given by $\tilde{D}_{\mu}\psi=\partial_{\mu}\psi+(1/2)\tilde{w}^{ab}_{\;\;\;\mu}\sigma_{ab}\psi$ and $\tilde{D}_{\mu}\bar{\psi}=\partial_{\mu}\bar{\psi}-(1/2)\tilde{w}^{ab}_{\;\;\;\mu}\bar{\psi}\sigma_{ab}$, where $\tilde{w}^{ab}_{\;\;\;\mu}=-\tilde{w}^{ba}_{\;\;\;\mu}$ are the spin connection components related to the Levi-Civita connection (also known as Ricci rotation coefficients), $w^{ab}_{\;\;\;\mu}=\tilde{w}^{ab}_{\;\;\;\mu}+K^{ab}_{\;\;\;\mu}$ is the RC spin connection and $\sigma_{ab}\equiv (1/2)\gamma^{[a}\gamma^{b]}$ are the generators of the Lorentz group in the spinor representation (GL(2,C)).}.
The Lagrangian can be expressed as
\begin{equation}
\mathcal{L}_{\rm Dirac}=\tilde{\mathcal{L}}_{\rm Dirac}+\dfrac{i\hbar}{8}K_{\alpha\beta\mu}\bar{\psi}\lbrace\gamma^{\mu},\gamma^{\alpha}\gamma^{\beta}\rbrace\psi \,,
\end{equation}
where $\tilde{\mathcal{L}}_{\rm Dirac}=\dfrac{i\hbar}{2}\left(\bar{\psi}\gamma^{\mu}\tilde{D}_{\mu}\psi-(\tilde{D}_{\mu}\bar{\psi})\gamma^{\mu}\psi\right)-m\bar{\psi}\psi \ $ and $\lbrace\gamma^{\mu},\gamma^{\alpha}\gamma^{\beta}\rbrace=\gamma^{\mu}\gamma^{\alpha}\gamma^{\beta}+\gamma^{\alpha}\gamma^{\beta}\gamma^{\mu}$.
Given the definition of the spin tensor in (\ref{spintensor}), we have
\begin{equation}
\mathcal{L}_{\rm Dirac}=\tilde{\mathcal{L}}_{\rm Dirac}+K^{\alpha\lambda\beta}s_{\lambda\beta\alpha}.
\end{equation}
This expression (as the previous one) is valid for any Dirac field minimally coupled to the RC spacetime geometry, it does not depend on any particular theory of gravity.

For this matter source the spin tensor is totally antisymmetric, i.e.,\footnote{Here we used the identities $\lbrace\gamma^{\mu},\gamma^{\alpha}\gamma^{\beta}\rbrace=2\gamma^{[\mu}\gamma^{\alpha}\gamma^{\beta]}=-2i\epsilon^{\mu\alpha\beta\lambda}\gamma_{\lambda}\gamma^{5}.$}
\begin{equation}
s^{\mu\nu\varepsilon}=\dfrac{1}{2}\epsilon^{\mu\nu\varepsilon\alpha}\breve{s}_{\alpha},
\end{equation}
and is expressed in terms of the Dirac (axial) spin vector as (see e.g. \cite{Obukhov,Poplawski:2011jz,Lucat:2015rla,Khanapurkar:2018gyo,Khanapurkar:2018jvx})
\begin{equation}
\breve{s}^{\beta}=\dfrac{\hbar}{2}\bar{\psi}\gamma^{\beta}\gamma^{5}\psi .
\end{equation}
This Dirac pseudo-vector field will play a crucial role later. Accordingly, the Cartan equations simplify since torsion is completely antisymmetric and Eq. (\ref{cartaneqs}) becomes
\begin{equation}
\label{contortiondirac}
T_{\alpha\beta\gamma}=K_{\alpha\beta\gamma}=\dfrac{\kappa^2}{2}\epsilon_{\alpha\beta\gamma\lambda}\breve{s}^{\lambda},
\end{equation}
therefore, the above Lagrangian introduces an effective spin-spin interaction induced by torsion.

Using the Cartan equations we then have
\begin{equation}
\mathcal{L}_{\rm Dirac}=\tilde{\mathcal{L}}_{\rm Dirac}+\kappa^{2}s^{\alpha\lambda\beta}s_{\lambda\beta\alpha}=\tilde{\mathcal{L}}_{\rm Dirac}-
\dfrac{3\kappa^{2}}{2}\breve{s}^{\lambda}\breve{s}_{\lambda}.
\end{equation}
The dynamical stress-energy tensor of Dirac fermions in EC theory is then given by
\begin{equation}
\label{stressenergyDirac}
T_{\mu\nu}=\tilde{T}^{\rm{Dirac}}_{\mu\nu}+\kappa^2\left(\dfrac{3}{2}\breve{s}^{\lambda}\breve{s}_{\lambda}g_{\mu\nu}-6\breve{s}_{\mu}\breve{s}_{\nu}\right),
\end{equation}
with $\tilde{T}^{\rm{Dirac}}_{\mu\nu}\equiv \dfrac{2}{\sqrt{-g}}\dfrac{\delta (\tilde{\mathcal{L}}_{\rm Dirac}\sqrt{-g})}{\delta g^{\mu}}$.
Moreover, we can also compute the form of the torsion-induced corrections on the right-hand side of the Einstein equations (\ref{ECeqs})
\begin{equation}
U_{\mu\nu}=\dfrac{2}{\sqrt{-g}}\dfrac{\delta (C\sqrt{-g})}{\delta g^{\mu}}=2\dfrac{\delta C}{\delta g^{\mu\nu}}-Cg_{\mu\nu},
\end{equation}
where in this case $C$ becomes simplified
\begin{eqnarray}
C&\equiv &-\dfrac{1}{2\kappa^2}
\Big(K^{\alpha\lambda}_{\;\;\;\alpha}K_{\gamma\lambda}^{\;\;\;\gamma}+K^{\alpha\lambda\beta}K_{\lambda\beta\alpha}\Big)
    \nonumber  \\
&=&-\dfrac{1}{2\kappa^2}
K^{\alpha\lambda\beta}K_{\lambda\beta\alpha}
    \nonumber  \\
&=&-\dfrac{\kappa^2}{2}
s^{\alpha\lambda\beta}s_{\lambda\beta\alpha}, 
\end{eqnarray}
due to the fact that for Dirac spinors contortion is completely antisymmetric.
Using the expressions above, the EC field equations (\ref{cartaneqs})--(\ref{ECeqs}) become %\footnote{NOTE FOR AUTHORS:
%Actually, if we use (20) and (23) we have $U_{\mu\nu}=\kappa^2\left(3\breve{s}_{\mu}
%\breve{s}_{\nu}-\dfrac{3}{4}\breve{s}^{\lambda}\breve{s}_{\lambda}g_{\mu\nu}\right)$ and 
%therefore we would get $\tilde{G}_{\mu\nu}=\kappa^2\tilde{T}^{\rm{Dirac}}_{\mu\nu}+
%\dfrac{3\kappa^4}{4}\breve{s}^{\lambda}\breve{s}_{\lambda}g_{\mu\nu}-3\kappa^{2}\breve{s}%_{\mu}\breve{s}_{\nu}$}
\begin{equation} \label{ECEinstein}
\tilde{G}_{\mu\nu} = \kappa^2\tilde{T}^{\rm{Dirac}}_{\mu\nu}+\dfrac{3\kappa^4}{4}\breve{s}^{\lambda}\breve{s}_{\lambda}g_{\mu\nu} ,
\end{equation}
and
\begin{equation}
T_{\alpha\beta\gamma}=K_{\alpha\beta\gamma}=\dfrac{\kappa^2}{2}\epsilon_{\lambda\alpha\beta\gamma}\breve{s}^{\lambda},
\end{equation}
respectively. Cosmological solutions in EC-Dirac theory have been investigated in detail in the literature, see e.g. \cite{Blagojevic:2013xpa,Obukhov, Lucat:2015rla,Poplawski:2011jz,Bronnikov:2016xvj}.

%%%%%%%%%%%%%%%%%%%%%%%%%%%%%%%%%%%%%%%%%%%%%%%%%%%%%%%%%%%%%%%%%%%%%%%%
\section{Einstein-Cartan-Dirac-Maxwell theory}\label{secIII}
%%%%%%%%%%%%%%%%%%%%%%%%%%%%%%%%%%%%%%%%%%%%%%%%%%%%%%%%%%%%%%%%%%%%%%%%

Let us now generalize the action (\ref{eq:actionEC}) to incorporate a minimal coupling between torsion and the electromagnetic field (for recent works on torsion-matter couplings see e.g. \cite{Harko:2014sja,Harko:2014aja,Carloni:2015lsa,Gonzalez-Espinoza:2018gyl,Jawad:2016zwj}). This can be directly implemented at the level of the matter fields by assuming the matter Lagrangian density 
\begin{equation}
\label{generalmatlagrange}
{L}_m=\mathcal{L}_{\rm D}+\mathcal{L}_{\rm M}+j^{\mu}A_{\mu},
\end{equation}
where the Dirac Lagrangian is the same as in Eq. (\ref{eq:mattfer}) therefore including a minimal coupling to the RC geometry, while $A_{\mu}$ is the electromagnetic four-potential and $j^{\nu}$ the electric charge current density of fermions. On the other hand, we now have the generalized Maxwell Lagrangian in a RC space-time (satisfying local Poincar\'{e} invariance) written as
\begin{equation}\label{eq:Maxfull}
\mathcal{L}_{\rm Max}=\frac{\lambda}{4}F_{\mu\nu}F^{\mu\nu} \ ,
\end{equation}
where $\lambda$ is a coupling parameter setting the system of units, and the generalized field strength tensor is defined as (note that $K^{\lambda}_{\;\;[\mu\nu]}=T^{\lambda}_{\;\;\mu\nu}$)
\begin{equation}
\label{newFaraday}
F_{\mu\nu}\equiv \nabla_{\mu}A_{\nu}-\nabla_{\nu}A_{\mu}=\tilde{F}_{\mu\nu}+2K^{\lambda}_{\;\;[\mu\nu]}A_{\lambda} \,,
\end{equation}
where $\nabla$ is the covariant derivative in RC space-time constructed with the independent connection $\Gamma_{\mu\nu}^{\lambda}$ in Eq. (\ref{eq:affcons}), while $\tilde{F}_{\mu\nu}=\partial_{\mu}A_{\nu}-\partial_{\nu}A_{\mu}$ is the standard field strength tensor when torsion is neglected. This expression of the electromagnetic field puts forward that, due to the presence of torsion in the minimal coupling, the second term in Eq.  (\ref{newFaraday}) breaks the $U(1)$ local symmetry. More explicitly, the Lagrangian density in Eq.  (\ref{eq:Maxfull}) becomes
\begin{equation}
\label{newMaxLagra}
\mathcal{L}_{\rm Max}=\tilde{\mathcal{L}}_{\rm Max}+\lambda \left(K^{\lambda[\mu\nu]}K^{\gamma}_{\;\;[\mu\nu]}A_{\gamma}+K^{\lambda[\mu\nu]}\tilde{F}_{\mu\nu} \right)A_{\lambda}.
\end{equation}

We will next proceed with the derivation of the gravitational, electromagnetic and fermionic field equations corresponding to the action (\ref{eq:actionEC}) with the $U(1)$-breaking term just introduced.

%%%%%%%%%%%%%%%%%%%%%%%%%%%%%%%%%%%%%%%%%%%%%%%%%%%%%%%%%%%%%%%%%%%%%%%%
\subsection{Gravitational sector}
%%%%%%%%%%%%%%%%%%%%%%%%%%%%%%%%%%%%%%%%%%%%%%%%%%%%%%%%%%%%%%%%%%%%%%%%
\subsubsection{Fermionic background torsion}

Let us assume a background torsion resulting from the spin density of fermionic fields. Variation of the action (\ref{eq:actionEC}) with respect to the metric for the above matter sources yields the gravitational equations in this case, which can be suitably written as
\begin{equation} \label{eq:ECDEgrav}
\tilde{G}_{\mu\nu}=\kappa^2\tilde{T}_{\mu\nu}+\dfrac{3\kappa^4}{4}\breve{s}^{\lambda}\breve{s}_{\lambda}g_{\mu\nu}+\kappa^{2}\Pi^{\rm int}_{\mu\nu}, 
\end{equation}
with 
\begin{equation}
\tilde{T}_{\mu\nu}=\tilde{T}^{\rm{Dirac}}_{\mu\nu}+\tilde{T}^{\rm{Max}}_{\mu\nu},
\end{equation}
and the new term (compare to Eqs. (\ref{ECEinstein})) induced by the minimal and non-minimal interactions between fermions and electromagnetic fields takes the form
\begin{equation}
\label{newinteractioninGR}
\Pi^{\rm int}_{\mu\nu}=\Pi^{U(1){\rm break}}_{\mu\nu}+4j_{(\mu}A_{\nu)}-j^{\lambda}A_{\lambda}g_{\mu\nu},
\end{equation}
where
\begin{eqnarray}
\label{energymomcorrec}
\Pi^{U(1){\rm break}}_{\mu\nu}&=&2\lambda \Big( T^{\;\;\alpha\beta}_{\mu}T^{\gamma}_{\;\;\alpha\beta}A_{\gamma}A_{\nu}+T^{\lambda\;\;\;\beta}_{\;\;\mu}T^{\gamma}_{\;\;\;\nu\beta}A_{\gamma}A_{\lambda}
	\nonumber \\
&&+ T^{\lambda\alpha}_{\;\;\;\;\mu}T^{\gamma}_{\;\;\alpha\nu}A_{\gamma}A_{\lambda}+T^{\lambda\alpha\beta}T_{\mu\alpha\beta}A_{\nu}A_{\lambda}
	\nonumber \\
&&+ T^{\;\;\alpha\beta}_{\mu}\tilde{F}_{\alpha\beta}A_{\nu}+T^{\lambda\;\;\;\beta}_{\;\;\mu}\tilde{F}_{\nu\beta}A_{\lambda}+T^{\lambda\alpha}_{\;\;\;\;\mu}\tilde{F}_{\alpha\nu}A_{\lambda}
	\nonumber \\
&&+\mu\leftrightarrow\nu \Big) \\
&&-\lambda\left(T^{\lambda\alpha\beta}T^{\gamma}_{\;\;\alpha\beta}A_{\gamma}A_{\lambda}+T^{\lambda\alpha\beta}\tilde{F}_{\alpha\beta}A_{\lambda}\right)g_{\mu\nu}\,. \nonumber
\end{eqnarray}
It is the (minimal) coupling of torsion with fermions and bosons that give rise to the non-minimal interactions between the matter fields, once the Cartan equations are used to replace the torsion components by the matter field variables. Under the ansatz that torsion is exclusively resulting from matter fields with half-integer intrinsic spin (fermions), it becomes completely antisymmetric and the Cartan equations in this case are still given by Eq.(\ref{contortiondirac}). This assumption is already   implicit in the second term of Einstein's equations above. This choice corresponds to keeping the (pseudo-trace) axial vector part of torsion as the only non-vanishing components.

Under such an ansatz, the interaction part (\ref{newinteractioninGR}) in the effective stress-energy tensor introduces terms both linear and quadratic in the spin density, all of which depend on the electromagnetic quantities. Accordingly, the value of the coupling constant $\lambda$ determines the scale at which the electromagnetic contribution becomes non-negligible. Thus, in principle, one could test torsion effects at spin densities smaller than the Cartan one, for sufficiently high electromagnetic fields, which suggests that new gravitational (metric) effects could be present in the core of magnetars \cite{Kaspi:2017fwg} and (hypothetical) quark stars \cite{Alford:2006vz}. Significant effects are expected for polarized matter because the linear terms will not average to zero and can introduce stronger torsion (spin) contributions to the Einstein equations at lower densities given a sufficiently high electromagnetic potential. In summary, torsion contributions to the metric field equations scale with $\kappa^{4}s^{2}$ for the pure EC spin density correction, and $\kappa^{4}\lambda s\tilde{F}A$ and $\kappa^{6}\lambda s^{2}A^{2}$ for the linear and quadratic $U(1)$ symmetry breaking terms, respectively.

In general, the  macroscopic description of a physical system is achieved through an averaging procedure. For simplicity, we will consider physical systems where the spin density obeys an approximate random distribution. This simplification (which is not valid in the presence of sufficiently intense magnetic fields that tend to align the spins) allows us to neglect the terms linear in the spin density and consider only the quadratic ones. For Dirac fermions, if we consider only the terms quadratic in torsion, we obtain after some algebraic manipulations
\begin{eqnarray}
\label{energymomcorr2}
\Pi^{U(1){\rm break}}_{\mu\nu}&=&\lambda\kappa^{4}\Bigg[\dfrac{g_{\mu\nu}}{2}\Big(A^{2}\breve{s}^{2}-(\breve{s}\cdot A)^{2}\Big)
\nonumber \\
&&-\breve{s}^{2}A_{\mu}A_{\nu}-A^{2}\breve{s}_{\mu}\breve{s}_{\nu}+4(\breve{s}\cdot A)\breve{s}_{(\mu}A_{\nu)}\Big].
\end{eqnarray}
Here, as usual $\breve{s}^{2}\equiv \breve{s}^{\lambda}\breve{s}_{\lambda}$, $A^{2}\equiv A^{\lambda}A_{\lambda}$ and $\breve{s}\cdot A\equiv \breve{s}^{\lambda}A_{\lambda}$.

\subsubsection{Full approach including the spin contribution from the electromagnetic sector }

If we also consider, besides the fermionic spin, the contribution from the generalized electromagnetic Lagrangian (\ref{newMaxLagra}) to the  (total) spin tensor, $s_{\lambda\alpha\beta}=s^{M}_{\lambda\alpha\beta}+s^{D}_{\lambda\alpha\beta}$, we obtain
\begin{equation}
s_{\lambda\alpha\beta}=\lambda\Big(A_{[\alpha}\tilde{F}_{\beta]\lambda}+2A_{[\alpha}T^{\gamma}{}_{\beta]\lambda}A_{\gamma}\Big)+s^{D}_{\lambda\alpha\beta},
\end{equation}
where $s_{\;\;\;\;\beta\gamma}^{D\;\alpha}$ is Dirac's spin tensor and 
\begin{eqnarray}
s_{\lambda\mu\nu}^{M}&=&\lambda A_{[\mu}F_{\nu]\lambda}
\nonumber \\
&=&\lambda\left(A_{[\mu}\tilde{F}_{\nu]\lambda}+2A_{[\mu}T^{\alpha}{}_{\;\nu]\lambda}A_{\alpha} \right),
\end{eqnarray}
represents the electromagnetic contribution to the spin tensor, i.e, 
%$s_{\lambda\mu\nu}^{M}=\dfrac{\delta \mathcal{L}^{U(1)break}_{M}}{\delta K^{\mu\nu\lambda}}$, 
$s_{\lambda\mu\nu}^{M}=\delta \mathcal{L}^{U(1)break}_{M}/\delta K^{\mu\nu\lambda}$,
which also depends on torsion due to the minimal coupling previously introduced.
The new Cartan equations can be written as
\begin{equation}
\label{torsionextended}
T^{\alpha}{}_{\beta\gamma}=\kappa^2\Big( s_{\;\;\;\;\beta\gamma}^{D\;\alpha}+ s_{\;\;\;\;\beta\gamma}^{M\;\alpha}+\delta^{\lambda}_{[\beta}s^{M}_{\gamma]}\Big),
\end{equation}
since Dirac's (completely antisymmetric) spin tensor
has zero trace vector.
With a bit of algebra we get
\begin{eqnarray}
\label{extendedCartan}
\tau^{\alpha}{}_{\beta\gamma}-2\lambda\kappa^{2} A_{\lambda}A_{[\beta}T^{\lambda\;\;\alpha}_{\;\;\gamma]}&=&\kappa^{2}\left(\dfrac{1}{2}\epsilon^{\alpha}{}_{\beta\gamma\lambda}\breve{s}^{\lambda}+\lambda A_{[\beta}\tilde{F}_{\gamma]}^{\;\;\alpha} \right), \nonumber
\end{eqnarray}
where $\tau^{\alpha}{}_{\beta\gamma}\equiv T^{\alpha}{}_{\beta\gamma}+T_{\gamma}\delta^{\alpha}_{\beta}-T_{\beta}\delta^{\alpha}_{\gamma}$ is the modified torsion tensor
and the first term on the right-hand side corresponds to Dirac's spin tensor previously introduced.
These expressions show that it is not trivial to separate the purely geometric torsion functions from the matter fields. 

Let us contract the indices $\alpha$ and $\gamma$ to obtain
\begin{equation}
-2T_{\beta}+\lambda\kappa^{2}A_{\lambda}T^{\lambda}_{\;\;\beta\gamma}A^{\gamma}=-\dfrac{\lambda\kappa^{2}}{2}\tilde{F}_{\beta\gamma}A^{\gamma}.
\end{equation}
which we shall use to find an expression for the torsion trace vector. From the equation above 
one gets the result $T_{\beta}A^{\beta}=0$ and this can  be used after contracting Eq.
 (\ref{extendedCartan}) with $A_{\alpha}A^{\gamma}$ to arrive at 
\begin{equation}
-A^{2}T_{\beta}+(1+\lambda\kappa^{2}A^{2})A_{\lambda}T^{\lambda}_{\;\;\beta\gamma}A^{\gamma}=-\dfrac{\lambda\kappa^{2}}{2}A^{2}\tilde{F}_{\beta\gamma}A^{\gamma}.
\end{equation}
Therefore, from this system of two equations we easily solve for $T_{\beta}$
\begin{equation}
T_{\beta}=-\dfrac{\lambda\kappa^{2}}{2(2+\lambda\kappa^{2}A^{2})}\tilde{F}_{\beta\gamma}A^{\gamma}.
\end{equation}
Proceeding in a similar manner, by contracting Eq. (\ref{extendedCartan}) with $A_{\alpha}$, after some algebra it is finally possible to transform the Cartan equations into a form in which the geometric torsion is separated from the matter fields, that is
\begin{eqnarray}
T^{\alpha}{}_{\beta\gamma}
&=&\kappa^{2}(\tilde{s}^{M\alpha}{}_{\beta\gamma}+s^{D\alpha}{}_{\beta\gamma}-2\lambda\kappa^{2}A_{\lambda}A_{[\beta}s^{D\lambda\alpha}{}_{\gamma}
	\nonumber \\
&&+\dfrac{2}{2+\lambda\kappa^{2}A^{2}}(\delta^{\alpha}_{[\beta}\tilde{s}^{M}_{\gamma]}-\lambda\kappa^{2}A^{\alpha}A_{[\beta}\tilde{s}^{M}_{\gamma]})),
\end{eqnarray}
which can be further simplified down to
\begin{eqnarray}
\label{newCartan}
T^{\alpha}{}_{\beta\gamma}
&=&\kappa^{2}(\tilde{s}^{M\alpha}{}_{\beta\gamma}+\rho^{\sigma}_{\beta}\rho^{\rho}_{\gamma}s^{D\alpha}{}_{\sigma\rho}
	\nonumber \\
&&+\dfrac{2}{2+\lambda\kappa^{2}A^{2}}(\delta^{\alpha}_{[\beta}\tilde{s}^{M}_{\gamma]}-\lambda\kappa^{2}A^{\alpha}A_{[\beta}\tilde{s}^{M}_{\gamma]})),
\end{eqnarray}
with 
\begin{equation}
\rho^{\alpha}_{\beta}\equiv \delta^{\alpha}_{\beta}+\lambda\kappa^{2}A^{\alpha}A_{\beta}\,,
\end{equation}
and we denote 
\begin{equation}
\tilde{s}^{M\alpha}{}_{\beta\gamma}\equiv\lambda A_{[\beta}\tilde{F}_{\gamma]}^{\;\;\alpha}\,,
\end{equation}
the torsion-free part of the (generalized) Maxwell spin tensor (therefore $\tilde{s}^{M}_{\beta}=-\dfrac{\lambda}{2}\tilde{F}_{\beta\gamma}A^{\gamma}$). A further simplification turns the Cartan equations into the final form
%\footnote{It is easy to verify that the product $\lambda\kappa^{2}A^{2}$ is dimensionless.}
\begin{eqnarray}
\label{finalnewCartan}
T^{\alpha}{}_{\beta\gamma}
&=&\kappa^{2}(\tilde{s}^{M\alpha}{}_{\beta\gamma}+
s^{D\alpha}{}_{\beta\gamma}+2\lambda\kappa^{2}s^{D\alpha\;\;\;\;\rho}_{\;\;\;[\beta}A_{\gamma]}A_{\rho}
	\nonumber \\
&&+\dfrac{2}{2+\lambda\kappa^{2}A^{2}}(\delta^{\alpha}_{[\beta}\tilde{s}^{M}_{\gamma]}-\lambda\kappa^{2}A^{\alpha}A_{[\beta}\tilde{s}^{M}_{\gamma]})),
\end{eqnarray}

This expression for torsion as a function of the matter fields can then be replaced in the matter Lagrangian (\ref{generalmatlagrange}), i.e, in the bosonic sector,
\begin{equation}
\label{BosonicLag}
\mathcal{L}_{\rm Max}=\tilde{\mathcal{L}}_{\rm Max}+\lambda \left(T^{\lambda\mu\nu}T^{\gamma}_{\;\;\mu\nu}A_{\gamma}+T^{\lambda\mu\nu}\tilde{F}_{\mu\nu} \right)A_{\lambda},
\end{equation}
and in the Dirac Lagrangian, 
\begin{equation}
\mathcal{L}_{\rm Dirac}=\tilde{\mathcal{L}}_{\rm Dirac}+
\dfrac{i\hbar}{4}K_{\alpha\beta\mu}\bar{\psi}\gamma^{[\mu}\gamma^{\alpha}\gamma^{\beta]}\psi,
\end{equation}
which after some algebra can be written as
\begin{equation}
\mathcal{L}_{\rm Dirac}=\tilde{\mathcal{L}}_{\rm Dirac}+3\breve{T}^{\lambda}\breve{s}^{Dirac}_{\lambda},
\end{equation}
 meaning that Dirac fields only interact with the axial vector part of torsion, $\breve{T}^{\lambda}\equiv \dfrac{1}{6}\epsilon^{\lambda\alpha\beta\gamma}T_{\alpha\beta\gamma}$. This expression is valid for any Dirac field minimaly coupled to a RC spacetime geometry, regardless of the gravitational theory.
  
Since bosons are also contributing to the torsion with the corresponding spin tensor, the axial torsion vector $\breve{T}^{\lambda}$ has now a new contribution (besides that of the Dirac axial spin vector).
From Eq. (\ref{newCartan}), we obtain
\begin{equation}
\breve{T}^{\lambda}=\kappa^{2}\left[-\dfrac{\breve{s}^{\lambda}}{2}+\dfrac{\lambda}{6}\epsilon^{\mu\beta\gamma\lambda}\left(2\kappa^{2}s^{D}_{\rho[\mu\beta}A_{\gamma]}A^{\rho}+A_{[\mu}\tilde{F}_{\beta\gamma]}\right)\right] \,,
\end{equation}
where we have dropped the $D$ symbol in the Dirac axial spin vector. Substituting in the Dirac Lagrangian above we get, after some algebra
\begin{eqnarray}
\label{newDiracLag}
\mathcal{L}_{\rm D}&=&\tilde{\mathcal{L}}_{\rm D}-\breve{s}_{\lambda}\breve{s}^{\lambda}\left(\dfrac{3\kappa^{2}}{2}+\lambda\kappa^{4}A^{2} \right)+\lambda\kappa^{4}(A\cdot \breve{s})^{2} 
\nonumber \\
&&+\dfrac{\lambda\kappa^{2}}{2}\epsilon^{\mu\beta\gamma\lambda}\breve{s}_{\lambda}A_{[\mu}\tilde{F}_{\beta\gamma]}.
\end{eqnarray}
The first term is Dirac's Lagrangian on a (pseudo) Riemann space-time, while the other terms come from the corrections of a RC geometry where torsion (given by the Cartan equations) is due to the spin tensors of fermionic spinors and electromagnetic fields.
The first term inside the parenthesis corresponds to  the well known spin-spin (axial-axial) contact interaction. Due to the presence of new fermionic-electromagnetic interactions induced by torsion, we see that the spin-spin contact interaction is now modulated at each point by the strength of the electromagnetic 4-potential (squared). The spin-spin effect is therefore affected locally by the electromagnetic potential at very high densities and fields, due to the $\kappa^{4}$ factor. The other two terms represent further (fermionic) spin-electromagnetic interactions. In the first of these, significant at very high densities and fields, the relative orientation (alignment) between the spin vector and the electromagnetic potential is relevant, which might suggest that this interaction could involve precession effects and possibly generate anisotropies in the spin distribution, for example via a macroscopic  (averaged) alignment of the fermionic spin.

To proceed with the analysis of the gravitational sector, we need to compute the effective stress-energy tensor $T^{\rm eff}_{\mu\nu}=T_{\mu\nu}+U_{\mu\nu}$. The dynamical stress-energy tensor $T_{\mu\nu}=\frac{2}{\sqrt{-g}} \frac{\delta (\sqrt{-g}\mathcal{L}_m)}{\delta g^{\mu\nu}}$ can be computed from the matter Lagrangian (\ref{generalmatlagrange}), using Eqs. (\ref{BosonicLag}) and (\ref{newDiracLag}), with the torsion corrections sourced by the new Cartan equations (\ref{newCartan}). These torsion-induced corrections to the stress-energy tensor $T_{\mu\nu}$ correspond to non-minimal interactions and also self-interactions of the matter fields. We can now write
\begin{equation}
T_{\mu\nu}=\tilde{T}_{\mu\nu}+4j_{(\mu}A_{\nu)}-j^{\lambda}A_{\lambda}g_{\mu\nu}+\Pi^{M{\rm int}}_{\mu\nu}+\Xi^{D{\rm int}}_{\mu\nu} \ ,
\end{equation}
where
 $\tilde{T}_{\mu\nu}= \tilde{T}^{\rm{Dirac}}_{\mu\nu}+\tilde{T}^{\rm{Max}} _{\mu\nu}$ is the standard stress-energy tensor for the matter-fields in curved space-time (that is, the one of GR),
$\Pi^{M{\rm int}}_{\mu\nu}$
comes from the bosonic Lagrangian (\ref{newMaxLagra}) and
includes non-minimal boson-fermion interactions (induced by torsion) and also bosonic self-interactions. Its explicit expression can be derived from Eq. (\ref{BosonicLag}) using the Cartan equations (\ref{newCartan}), and gives rise to a quite long expression containing many terms (non-minimal and self-interactions). If we consider the regime of approximate random distributions of fermionic spin, corresponding to unpolarized matter (in the sense of zero macroscopic intrinsic magnetic moment), we end up with a simplified expression, since all terms that scale linearly with the fermionic spin are neglected. On the other hand, $\Xi^{D{\rm int}}_{\mu\nu}$ comes from the Dirac Lagrangian, i.e, $\Xi^{D{\rm int}}_{\mu\nu}=\dfrac{2}{\sqrt{-g}}\dfrac{\partial L^{Corr}_{Dirac}}{\partial g^{\mu\nu}}$, where $L^{Corr}_{Dirac}\equiv3\breve{T}^{\lambda}\breve{s}^{Dirac}_{\lambda}$, and it corresponds to non-minimal fermion-boson interactions (induced by torsion) and also spin-spin fermionic self-interactions. This tensor can be computed exactly as
\begin{eqnarray}
\label{interactDirac}
\Xi^{D{\rm int}}_{\mu\nu}&=&
-4\Bigg(\left(\dfrac{3\kappa^{2}}{2}+\lambda\kappa^{4}A^{2} \right)\breve{s}_{\mu}\breve{s}_{\nu}+\lambda\kappa^{4}\breve{s}^{2}A_{\mu}A_{\nu}\Bigg)
	\nonumber \\
&&+8\lambda\kappa^{4}(A\cdot\breve{s})A_{(\mu}\breve{s}_{\nu)}
	\nonumber\\
&&-\Bigg(\lambda\kappa^{4}(A\cdot\breve{s})^{2}-\breve{s}^{2}\left(\dfrac{3\kappa^{2}}{2}+\lambda\kappa^{4}A^{2} \right) 
\Bigg)g_{\mu\nu},\nonumber \\ 
\end{eqnarray}
assuming random fermionic spin distributions, therefore neglecting the last term in (\ref{newDiracLag}).

Now we can compute the correction to the bosonic Lagrangian as
\begin{eqnarray}
\label{BosonLagcorr}
\mathcal{L}^{M}_{\rm corr}&\approx&\lambda^{2}\kappa^{2}A^{[\mu}\tilde{F}^{\nu]\lambda}\tilde{F}_{\mu\nu}A_{\lambda}
\nonumber \\
&&+\dfrac{2\lambda\kappa^{2}\tilde{F}_{\mu\nu}}{2+\lambda\kappa^{2}A^{2}}A^{[\mu}\tilde{s}^{\nu]}(1-\lambda \kappa^{2}A^{2})\nonumber \\
&&+\lambda^{3}\kappa^{4}A^{[\mu}\tilde{F}^{\nu]\lambda}A_{[\mu}\tilde{F}_{\nu]\gamma}A_{\lambda}A^{\gamma}\nonumber \\
&&+\dfrac{4\lambda\kappa^{4}A^{[\mu}\tilde{s}^{\nu]}A_{[\mu}\tilde{s}_{\nu]}}{(2+\lambda\kappa^{2}A^{2})^{2}}(1-\lambda\kappa^{2}A^{2}(2-\lambda\kappa^{2}A^{2}))\nonumber \\
&&-\dfrac{\lambda\kappa^{4}}{2}(A^{2}\breve{s}^{2}-(A\cdot\breve{s})^{2}).
\end{eqnarray}
The last term here depends on the spinors via Dirac axial vector $\breve{s}^{\lambda}$ and represents non-minimal boson-fermion interactions, while every other term in that expression corresponds to self-interactions\footnote{The third and fourth terms in the Lagrangian above can be re-written as $\dfrac{\lambda^{3}}{2}\kappa^{4}A^{2}\tilde{F}^{\nu}_{\;\;\lambda}\tilde{F}_{\nu\gamma}A^{\lambda}A^{\gamma}$ and $\dfrac{2\lambda\kappa^{4}(\tilde{s}^{2}A^{2}-(A\cdot\tilde{s})^{2})}{(2+\lambda\kappa^{2}A^{2})^{2}}(1-\lambda\kappa^{2}A^{2}(2-\lambda\kappa^{2}A^{2}))$, respectively}. Note that we have dropped the M from the trace vector of what we called the torsionless part of the bosonic spin tensor $\tilde{s}^{M\alpha}{}_{\beta\gamma}\equiv\lambda A_{[\beta}\tilde{F}_{\gamma]}^{\;\;\alpha}$. A lengthy and tedious process allows to compute the total stress-energy contribution from non-minimal and self-interactions as
\begin{eqnarray}
\label{interactbosonic}
\Pi^{M{\rm int}}_{\mu\nu}+\Xi^{D{\rm int}}_{\mu\nu}&=&
2\lambda^{2}\kappa^{2}\tilde{F}_{\alpha\beta} A^{[\alpha}\tilde{F}^{\beta]}_{\;\;\mu}A_{\nu}+\dfrac{4\lambda\kappa^{2}}{2+\lambda\kappa^{2}A^{2}}\times
	\nonumber \\
&&\times\left(\tilde{s}^{\beta}\tilde{F}_{\mu\beta}-\lambda\kappa^{2}A^{[\alpha}\tilde{s}^{\beta]}\tilde{F}_{\alpha\beta}A_{\mu}\right)A_{\nu}
	\nonumber \\
&&+4\lambda^{3}\kappa^{4}A^{2}A_{\lambda}A_{[\mu}\tilde{F}_{\gamma]}^{\;\;\lambda}\tilde{F}_{\nu}^{\;\;\gamma}
	\nonumber \\
&&+\dfrac{4\lambda\kappa^{4}A_{[\mu}\tilde{s}_{\gamma]}\tilde{F}_{\nu}^{\;\;\gamma}}{2+\lambda\kappa^{2}A^{2}}(1-\lambda\kappa^{2}A^{2})
	\nonumber \\ 
&&+(\mu \leftrightarrow \nu)
	\nonumber \\ 
&&+2\lambda^{3}\kappa^{4}\Big(A^{2}\tilde{F}_{\mu}^{\;\;\lambda}\tilde{F}_{\nu}^{\;\;\gamma}+A_{\mu}A_{\nu}\tilde{F}_{\alpha}^{\;\;\lambda}\tilde{F}^{\alpha\gamma}\nonumber \\
&&-2A_{(\mu}\tilde{F}_{\nu)}^{\;\;\gamma}\tilde{F}_{\alpha}^{\;\;\lambda}A^{\alpha}\Big)A_{\lambda}A_{\gamma}\nonumber \\
&&+16\lambda\kappa^{4}A^{2}\tilde{F}^{\alpha}_{\;\;\mu}A_{\nu}\tilde{F}_{\alpha\gamma}A^{\gamma}+A_{\mu}A_{\nu}\times
	\nonumber \\
&&\times \left( \tilde{s}^{2}h(A)-(A^{2}\tilde{s}^{2}-(A\cdot \tilde{s})^{2})t(A)\right) 
	\nonumber \\ 
&&+(A^{2}\tilde{s}_{\mu}\tilde{s}_{\nu}-(A\cdot \tilde{s})\tilde{s}_{(\mu}A_{\nu)})h(A)
\nonumber \\
&&-\mathcal{L}^{M}_{\rm self}g_{\mu\nu} \nonumber \\
&&-6\left( \kappa^{2}+\lambda\kappa^{4}A^{2} \right)\breve{s}_{\mu}\breve{s}_{\nu}-6\lambda\kappa^{4}\breve{s}^{2}A_{\mu}A_{\nu} 
	\nonumber \\
&&+16\lambda\kappa^{4}(A\cdot\breve{s})A_{(\mu}\breve{s}_{\nu)}
-\dfrac{1}{2}\Big(\lambda\kappa^{4}(A\cdot\breve{s})^{2}
	\nonumber\\
&&-\breve{s}^{2}\left(3\kappa^{2}+\lambda\kappa^{4}A^{2} \right) 
\Big)g_{\mu\nu} \,,
\end{eqnarray} 
where the functions $h(A)$ and $t(A)$ are given by
\begin{eqnarray}
h(A)&=&\dfrac{4\lambda\kappa^{4}}{(2+\lambda\kappa^{2}A^{2})^{2}}(1-\lambda^{2}\kappa^{2}A^{2}(2-\kappa^{2}A^{2})),\nonumber \\
t(A)&=&\dfrac{8\lambda^{2}\kappa^{6}(\lambda\kappa^{2}A^{2}+1)}{(2+\lambda\kappa^{2}A^{2})^{2}}\nonumber \\
&&+\dfrac{16\lambda\kappa^{2}}{(2+\lambda\kappa^{2}A^{2})^{3}}(1-\lambda^{2}\kappa^{2}A^{2}(2-\lambda\kappa^{2}A^{2})) \nonumber \ ,
\end{eqnarray}
respectively. In the expression (\ref{interactbosonic}) only the last four terms depend on the spinors via the Dirac spin axial vector. The term $\mathcal{L}^{M}_{\rm self}g_{\mu\nu}$ corresponds to the purely electromagnetic terms of Eq.(\ref{BosonLagcorr}), i.e, the self-interactions. In absence of electromagnetic potentials we recover the second term of Eq.(\ref{stressenergyDirac}) of the EC-Dirac model. Let us illustrate these considerations with an example. Assuming homogeneity and isotropy (FRWL metric) and $A_{\mu}=(\phi(t),0,0,0)$, which means $\tilde{F}_{\mu\nu}=0$ and $\tilde{s}^{\alpha\beta\gamma}=0$, we obtain for the combination above
\begin{eqnarray}
\Pi^{M{\rm int}}_{\mu\nu}+\Xi^{D{\rm int}}_{\mu\nu}&=&
-6\left( \kappa^{2}+\lambda\kappa^{4}A^{2} \right)\breve{s}_{\mu}\breve{s}_{\nu}-6\lambda\kappa^{4}\breve{s}^{2}A_{\mu}A_{\nu} 
	\nonumber \\
&&+16\lambda\kappa^{4}(A\cdot\breve{s})A_{(\mu}\breve{s}_{\nu)}
-\dfrac{1}{2}\Big(\lambda\kappa^{4}(A\cdot\breve{s})^{2}
	\nonumber \\
&&-\breve{s}^{2}\left(3\kappa^{2}+\lambda\kappa^{4}A^{2} \right) 
\big)g_{\mu\nu} \,, 
\end{eqnarray}
whose components give the energy density 
\begin{eqnarray}
\Pi^{M{\rm int}}_{00}+\Xi^{D{\rm int}}_{00}&=&\breve{s}_{0}^{2}\left( \dfrac{19}{2}\lambda\kappa^{4}\phi^{2}-6\kappa^{2}\right) 
	\nonumber \\
&&+\breve{s}^{2}\left( \dfrac{3\kappa^{2}}{2}-\dfrac{11}{2}\lambda\kappa^{4}\phi^{2}\right) \ ,
\end{eqnarray}
and the pressure terms
\begin{eqnarray}
\Pi^{i}_{j}+\Xi^{i}_{j}&=&-\breve{s}^{i}\breve{s}_{j}\left(6\lambda\kappa^{4}\phi^{2}+6\kappa^{2}\right) \nonumber \\
&&-\left(\dfrac{\lambda\kappa^{4}}{2}\breve{s}_{0}^{2}\phi^{2}-\breve{s}^{2}(\dfrac{3\kappa^{2}}{2}+\dfrac{\lambda\kappa^{4}}{2}\phi^{2})\right)\delta^{i}_{j} \ ,
\nonumber \\
\end{eqnarray}
respectively.

Let us focus now our attention upon the geometrical term $U_{\mu\nu}=\frac{2}{\sqrt{-g}} \frac{\delta (\sqrt{-g}C)}{\delta g^{\mu\nu}}$, where
\begin{equation}
C=-\dfrac{\kappa^{2}}{2}\Big(s^{\lambda}s_{\lambda}+ s^{\mu\nu\lambda}\left(s_{\nu\lambda\mu}+s_{\lambda\mu\nu}+s_{\mu\lambda\nu}\right)\Big) \ .
\end{equation}
With a bit of algebra we obtain
\begin{eqnarray}
C&=&-\dfrac{\kappa^{2}}{2}\Big(s^{M}_{\gamma}s^{M\;\gamma}+s^{M\;\mu\nu\lambda}\left(s^{M}_{\mu\lambda\nu}+2s^{M}_{[\nu\lambda]\mu}\right)
\nonumber \\
&&+2s^{M\;\mu\nu\lambda}s^{D}_{\nu\lambda\mu}-\dfrac{3}{2}\breve{s}^{D}_{\lambda}\breve{s}^{D\;\lambda}\Big).
\end{eqnarray}
Note that torsion appears in the first three terms, where it has to be replaced by the corresponding matter (spin) quantities via Cartan's Eq.(\ref{newCartan}). Formally, the tensor $U_{\mu\nu}$ obeys the following expression
\begin{eqnarray}
U_{\mu\nu}&=&
-\kappa^{2}\Big(2s^{M}_{\mu}s^{M}_{\nu}+s^{M\;\alpha\beta}_{\mu}(s^{M}_{\nu\beta\alpha}+2s^{M}_{[\alpha\beta]\nu}-2s^{D}_{\nu\beta\alpha})
	\nonumber\\
&&+s^{M\;\lambda\;\;\beta}_{\;\;\;\;\;\;\;\mu}(s^{M}_{\lambda\beta\nu}+2 s^{M}_{[\nu\beta]\lambda}-2s^{D}_{\lambda\beta\nu})
	\nonumber \\
&&+s^{M\;\lambda\alpha}_{\;\;\;\;\;\;\;\;\mu}(s^{M}_{\lambda\nu\alpha}+2s^{M}_{[\nu\lambda]\mu}-2s^{D}_{\lambda\nu\alpha})-3\breve{s}^{D}_{\mu}\breve{s}^{D}_{\nu}\Big)
	\nonumber \\
&&-Cg_{\mu\nu},
\end{eqnarray}
and the resulting expression, as a function of the matter fields, upon substitution of torsion via Cartan's equations in (\ref{finalnewCartan}) is rather long and complicated. To simplify things we will assume again spatial homogeneity and isotropy, which allows us to write the torsion tensor 
\begin{eqnarray}
T_{\alpha\beta\gamma}&=&\kappa^{2}\left(s^{D}_{\alpha\beta\gamma}+2\lambda\kappa^{2}s^{D\;\;\;\;\rho}_{\alpha[\beta}A_{\gamma]}A_{\rho} \right), 
\end{eqnarray}
therefore we find
\begin{equation}
s^{M}_{\mu}=\tilde{s}^{M}_{\mu}+\lambda T_{(\alpha\beta)\mu}A^{\alpha}A^{\beta}=0,
\end{equation} 
and
\begin{equation}
s^{M}_{\alpha\beta\gamma}=2\lambda\kappa^{2}A_{[\beta}s^{D\;\lambda}_{\;\;\;\;\;\;\gamma]\alpha}A_{\lambda}.
\end{equation}
Using these expressions, after some algebra we obtain
\begin{eqnarray}
C&=&-\dfrac{\kappa^{2}}{2}\Big(\lambda\kappa^{2}(2-\lambda\kappa^{2}A^{2})(A^{2}\breve{s}^{2}-(A\cdot\breve{s})^{2})-\dfrac{3}{2}\breve{s}^{2}\Big) \nonumber  \ ,
\end{eqnarray}
and the corresponding stress-energy tensor is 
\begin{eqnarray}
U_{\mu\nu}&=&-2\lambda\kappa^{4}\Big(A_{\mu}A_{\nu}(\breve{s}^{2}(2-\lambda\kappa^{2}A^{2})
	\nonumber \\
&&-\lambda\kappa^{2}(A^{2}\breve{s}^{2}-(A\cdot\breve{s})^{2}))
	\nonumber \\
&&+\breve{s}_{\mu}\breve{s}_{\nu}\left[A^{2}(2-\lambda\kappa^{2}A^{2})-\dfrac{3}{2\lambda\kappa^{2}}\right]  
	\nonumber \\
&&+2(2-\lambda\kappa^{2}A^{2})(A\cdot\breve{s})A_{(\mu}\breve{s}_{\nu)}\Big)
	\nonumber \\
&&-Cg_{\mu\nu} \nonumber \ .
\end{eqnarray}
This completes the calculations for the effective stress-energy tensor $T^{\rm eff}_{\mu\nu}=T_{\mu\nu}+U_{\mu\nu}$, that enters the right-hand side of the Einstein equations
\begin{equation}
\tilde{G}_{\mu\nu}=\kappa^{2}T^{\rm eff}_{\mu\nu}.
\end{equation}

As one can see from the expressions obtained in the right-hand side of this equation, if quarks can form a condensate in vacuum, corresponding to non-zero vacuum expectation values for the quadratic terms $\sim\breve{s}^{2}$ in the above equations, then the model predicts an effective cosmological constant modulated by the bosonic dynamics, providing a dynamical dark energy from vacuum condensates. This is possible thanks to the non-minimal and self-interactions involving the square of the Dirac (axial) spin.

The total matter Lagrangian will give rise to extended Dirac and electromagnetic equations. To compute this we will now analyze the bosonic and fermionic field dynamics.

%%%%%%%%%%%%%%%%%%%%%%%%%%%%%%%%%%%%%%%%%%%%%%%%%%%%%%%%%%%%%%%%%%%%%%%%
\subsection{Electromagnetic sector}
%%%%%%%%%%%%%%%%%%%%%%%%%%%%%%%%%%%%%%%%%%%%%%%%%%%%%%%%%%%%%%%%%%%%%%%%

Let us now obtain the electromagnetic field equations. Consider the action constructed from Eq. (\ref{eq:Maxfull}) together with the usual source term $j^{\lambda}A_{\lambda}$. Varying it with respect to the vector potential $A_{\mu}$ yields
\begin{equation}
\nabla_{\mu}F^{\mu\nu}=\lambda^{-1}j^{\nu} \ ,
\end{equation}
which can be conveniently rewritten as
\begin{equation} \label{eq:ECDEem}
\qquad \tilde{\nabla}_{\mu}\tilde{F}^{\mu\nu}=\lambda^{-1}(j^{\nu}+J^{\nu}) \ ,
\end{equation}
where we have defined the torsion-induced four-current
\begin{eqnarray}
\label{newcurrentgeneral}
J^{\nu}&=&-\lambda\Big[2(K^{\nu}_{\;\;\lambda\mu}K^{\gamma[\mu\lambda]}+K_{\lambda}K^{\gamma[\lambda\nu]})A_{\gamma}\,
	\nonumber \\
&&+K^{\nu}_{\;\;\lambda\mu}\tilde{F}^{\mu\lambda}+K_{\lambda}\tilde{F}^{\lambda\nu}
+2\tilde{\nabla}_{\mu}\left(K^{\gamma[\mu\nu]}A_{\gamma}\right)\Big] ,
\end{eqnarray}
with $K_{\lambda}\equiv K^{\alpha}_{\;\;\lambda\alpha}$. As can be seen in the expression for the Lagrangian in Eq. (\ref{newMaxLagra}), or in the field equations  (\ref{eq:ECDEem}), the terms quadratic in the contortion or, equivalently, in the spin density, resemble Proca-like terms. From this analogy, the coupling between the electromagnetic four-potential and the space-time torsion provides an effective mass for the photon $m_{\gamma}^{2}\sim \lambda T^{2}$ in physical environments where the $U(1)$-breaking phase transition takes place. The terms linear in torsion, on the other hand, reveal new physical effects due to the coupling between electromagnetism and torsion, which in this framework become significant for spin densities much lower than the Cartan density. That is, way before the manifestation of new metric effects that are implicit in Eqs. (\ref{eq:ECDEgrav}) and (\ref{newinteractioninGR}), the torsion (spin) of fermions start interacting significantly with electromagnetism, affecting  Maxwell's dynamics. This is another motivation to consider physical effects of the full dynamics in astrophysical and cosmological environments with spin densities below the Cartan threshold, as in the core of neutron stars and in the early universe.

To complete the electromagnetic sector of the dynamics, we include the generalized conservation equation, which can be written as
\begin{equation}
\label{chargeequation}
\tilde{\nabla}_{\nu}j^{\nu}=-\tilde{\nabla}_{\nu}J^{\nu}.
\end{equation}
Alternatively, one can write
\begin{equation}
\nabla_{\nu}j^{\nu}=\dfrac{\lambda}{2}\left[ \nabla_{\nu},\nabla_{\mu}\right] F^{\mu\nu} \ ,
\end{equation}
where
\begin{equation}
\left[\nabla_{\nu},\nabla_{\mu}\right] F^{\mu\nu}=R^{\mu}_{\;\;\varepsilon\nu\mu}F^{\varepsilon\nu}+R^{\nu}_{\;\;\varepsilon\nu\mu}F^{\mu\varepsilon}+2T^{\gamma}_{\;\;\nu\mu}\nabla_{\gamma}F^{\mu\nu},
\end{equation}
is the commutator of covariant derivation of an antisymmetric $(0,2)$-tensor in RC space-time. %This expression is valid for all gravity theories with this RC space-time geometry.
We see that Dirac's current is not conserved, therefore, from a probabilistic semi-quantum description point of view the particle number can change due to intense gravitational fields.

\subsubsection{Fermionic background torsion}

Assuming the ansatz of a completely antisymmetric background torsion, as in the case where torsion comes from the background Dirac fermionic fields, we get the same form of the field equations but the torsion-induced current gets simplified
\begin{equation}
\label{newcurrent}
J^{\nu}=-\lambda\left[2K^{\nu}_{\;\;\lambda\mu}K^{\gamma\mu\lambda}A_{\gamma}-K^{\nu}_{\;\;\lambda\mu}\tilde{F}^{\lambda\mu}+2\tilde{\nabla}_{\mu}(K^{\gamma\mu\nu}A_{\gamma})
\right].
\end{equation}
According to the minimal coupling between torsion and electromagnetic fields, as it is apparent from Eq. (\ref{newFaraday}), only the antisymmetric part of the contortion tensor enters the electromagnetic sector in a RC space-time (at the Lagrangian level). However, for fermions both torsion and contortion are totally antisymmetric, so we have dropped out the brackets for antisymmetrization. In that case it is useful to express the Maxwell Lagrangian with torsion contributions, Eq. (\ref{newMaxLagra}), as
\begin{equation}
\label{x}
\mathcal{L}_{\rm Max}=\tilde{\mathcal{L}}_{\rm Max}-\lambda\left[ \dfrac{\kappa^{4}}{2}\left(\breve{s}^{2}A^{2}-(\breve{s}\cdot A)^2\right)-\dfrac{\kappa^2}{2}f^{\nu}\breve{s}_{\nu}\right],
\end{equation}
where we have introduced the (axial) vector
\begin{equation}
\label{smallf}
f^{\rho}\equiv \epsilon^{\lambda\mu\nu\rho}A_{\lambda}\tilde{F}_{\mu\nu}.
\end{equation}
Under the assumption of the random spin distribution, from Eq. (\ref{newcurrent}) and using the Cartan equations (\ref{contortiondirac}), we obtain the following (spin) torsion-induced four-current 
\begin{eqnarray}
\label{torsioninducedcurrent}
J^{\nu}&=&-\kappa^{4}\lambda \left(\breve{s}^{2}A^{\nu}-(\breve{s}\cdot A)\breve{s}^{\nu}\right),
\end{eqnarray}
which arises from the interaction between the fermionic axial vector field and the electromagnetic 4-potential.
\subsubsection{Full approach including the contribution of the generalized electromagnetism to the spin tensor}

In this case, the Cartan equations are given by Eq. (\ref{newCartan}). We will consider for convenience the generalized current in the following form
\begin{eqnarray}
J^{\nu}&=&-\lambda\Big[2(T^{\nu}_{\;\;\lambda\mu}T^{\gamma\mu\lambda}+2T_{\lambda}T^{\gamma\lambda\nu})A_{\gamma}\,
\nonumber \\
&&+T^{\nu}_{\;\;\lambda\mu}\tilde{F}^{\mu\lambda}+2T_{\lambda}\tilde{F}^{\lambda\nu}
+2\tilde{\nabla}_{\mu}(T^{\gamma\mu\nu}A_{\gamma})\Big] \ ,
\end{eqnarray}
where we have used the fact that contortion is antisymmetric in the first two indices and also that $K^{\nu}_{\;\;[\lambda\mu]}=T^{\nu}_{\;\;\lambda\mu}$ and $K_{\lambda}=2T_{\lambda}$. Now, given the fact that the total matter Lagrangian can be written as
\begin{equation}
\mathcal{L}_{\rm mat}=\mathcal{L}_{\rm D}+\mathcal{L}_{\rm M}+j^{\mu}A_{\mu} \ ,
\end{equation}
where $\mathcal{L}_{\rm D}$ includes bosonic-fermionic interactions and is given by Eq. (\ref{newDiracLag}) and $\mathcal{L}_{\rm M}$ is given in Eq. (\ref{newMaxLagra}), upon applying the variational principle with respect to the electromagnetic potential, we get a new generalized Maxwell equation in Eq. (\ref{eq:ECDEem}) given by
\begin{equation} \label{newEMeq}
\qquad \tilde{\nabla}_{\mu}\tilde{F}^{\mu\nu}=\lambda^{-1}(j^{\nu}+J^{\nu}+\xi^{\nu}) \ ,
\end{equation}
where
\begin{eqnarray}
\xi^{\nu}&=&\lambda\Big(2\kappa^{4}\left[\breve{s}^{\lambda}(A\cdot \breve{s}) -\breve{s}_{\lambda}\breve{s}^{\lambda}A^{\nu} \right] 
	\nonumber \\
&&+\dfrac{\lambda\kappa^{2}}{2}\left[ \epsilon^{\nu\beta\gamma\lambda}\tilde{F}_{\beta\gamma}\breve{s}_{\lambda}-2\epsilon^{\rho\mu\nu\lambda}\tilde{\nabla}_{\mu}(A_{\rho}\breve{s}_{\lambda})\right] \Big) ,
\end{eqnarray}
comes from the (effective) Dirac Lagrangian (\ref{newDiracLag}) as 
\begin{equation}
\xi^{\nu}\equiv\dfrac{\partial\mathcal{L}_{\rm D}}{\partial A_{\nu}}-\tilde{\nabla}_{\mu}\left(\dfrac{\partial\mathcal{L}_{\rm D}}{\partial (\tilde{\nabla}_{\mu}A_{\nu})} \right) \,. \nonumber
\end{equation}
%
%$\xi^{\nu}\equiv\dfrac{\partial\mathcal{L}_{\rm D}}{\partial A_{\nu}}-\tilde{\nabla}_{\mu}\left(\dfrac{\partial
%\mathcal{L}_{\rm D}}{\partial (\tilde{\nabla}_{\mu}A_{\nu})} \right) $. 
%
Using now Eq. (\ref{newCartan}) we obtain a long expression for the torsion-induced current $J^{\nu}$ with non-linear terms. One can also use the effective Maxwell Lagrangian in Eq. (\ref{BosonLagcorr}) to obtain
\begin{equation}
J^{\nu} \equiv \dfrac{\partial\mathcal{L}^{corr}_{\rm M}}{\partial A_{\nu}}-\tilde{\nabla}_{\mu}\left(\dfrac{\partial\mathcal{L}^{corr}_{\rm M}}{\partial (\tilde{\nabla}_{\mu}A_{\nu})} \right) \,, \nonumber
\end{equation}
%
% $J^{\nu} \equiv \dfrac{\partial\mathcal{L}^{corr}_{\rm M}}{\partial A_{\nu}}-\tilde{\nabla}_{\mu}
%\left(\dfrac{\partial\mathcal{L}^{corr}_{\rm M}}{\partial (\tilde{\nabla}_{\mu}A_{\nu})} \right) $ 
as
\begin{eqnarray}
J^{\nu}&=&\lambda\kappa^{2}\Big[\tilde{F}_{\alpha\beta}\left(\lambda A^{[\alpha}\tilde{F}^{\beta]\nu}+2A^{[\alpha}\tilde{s}^{\beta]}A^{\nu}X(A)\right)\nonumber \\
&&+2\tilde{F}^{\nu}_{\;\;\beta}\left({F}^{\beta}_{\;\;\lambda}A^{\lambda} +2{s}^{\beta}Y(A)\right)\nonumber \\
&&+\lambda^{2}\kappa^{2}\left(A^{\nu}\tilde{F}^{\alpha\lambda}A_{\lambda}+A^{2}\tilde{F}^{\alpha\nu}\right)\tilde{F}_{\alpha\gamma}A^{\gamma}\nonumber \\
&&+(A^{\nu}\tilde{s}^{2}-2\tilde{s}^{\nu}(A\cdot\tilde{s}))Z(A) \nonumber \\
&&+A^{\nu}(A^{2}\tilde{s}^{2}-(A\cdot\tilde{s})^{2})W(A)\nonumber \\
&&-\kappa^{2}(A^{\nu}\breve{s}^{2}-\breve{s}^{\nu}(A\cdot\breve{s}))\Big]
\nonumber \\
&&-
\tilde{\nabla}_{\mu}\left(\dfrac{\partial\mathcal{L}^{corr}_{\rm M}}{\partial (\tilde{\nabla}_{\mu}A_{\nu})} \right)  \,,
\end{eqnarray}
where the last term is computed as
\begin{eqnarray}
\dfrac{\partial\mathcal{L}^{corr}_{\rm M}}{\partial (\tilde{\nabla}_{\mu}A_{\nu})}&=&2\lambda^{2}\kappa^{2}\Big(A^{[\mu}\tilde{F}^{\nu]\lambda}A_{\lambda}+\tilde{F}^{\alpha[\mu}A^{\nu]}A_{\alpha} 
	\nonumber \\
&&
-\tilde{F}^{[\mu}_{\;\;\beta}A^{\nu]}A^{\beta}\Big) 
+4\lambda\kappa^{2}A^{[\mu}\tilde{s}^{\nu]}\dfrac{1-\lambda\kappa^{2}A^{2}}{2+\lambda\kappa^{2}A^{2}}
	\nonumber \\  
&&+\lambda^{3}\kappa^{4}A^{2}\tilde{F}^{[\mu}_{\;\;\gamma}A^{\nu]}A^{\gamma}    \, ,
\end{eqnarray}
and we have introduced the definitions
\begin{eqnarray}
X(A)&\equiv & -\dfrac{6\lambda\kappa^{2}}{(2+\lambda\kappa^{2}A^{2})^{2}} \,, \nonumber \\
Y(A)&\equiv &-\dfrac{1-\lambda\kappa^{2}A^{2}}{2+\lambda\kappa^{2}A^{2}} \,, \nonumber \\
Z(A)&\equiv & \dfrac{2\kappa^{2}(1-(2-\lambda\kappa^{2}A^{2}))}{2+\lambda\kappa^{2}A^{2}} \,, \nonumber \\
W(A)&\equiv &\Big[4\lambda\kappa^{2}\Big((2+\lambda\kappa^{2}A^{2})(\lambda\kappa^{2}A^{2}-1) \,,
	\nonumber \\
&&-\left(1-\lambda\kappa^{2}A^{2}(2-\lambda\kappa^{2}A^{2})\right)\Big)\Big]/(2+\lambda\kappa^{2}A^{2})^{3} \nonumber \,. 
\end{eqnarray}
These highly involved expressions can be interpreted as non-linear electrodynamics with non-minimal couplings between fermionic matter (spinors) and electromagnetic fields induced by the RC space-time geometry. These equations are simplified in two cases: (i) matter with a random distribution of fermionic spins, where we neglect all quantities linear in the Dirac spin, leaving only the quadratic ones which do not vanish after macroscopic averaging and (ii) the case of homogeneity and isotropy, with $A=(\phi,0,0,0)$, and $\tilde{s}=0=\tilde{F}$. In the first case we obtain
\begin{eqnarray}
\xi^{\nu}&\approx&2\lambda\kappa^{4}\left[\breve{s}^{\nu}(A\cdot \breve{s}) -\breve{s}_{\lambda}\breve{s}^{\lambda}A^{\nu} \right],
\end{eqnarray}
and in the second case, the simplified $J^{\nu}$ is simply
\begin{eqnarray}
J^{\nu}&=&-\lambda\kappa^{4}
\left[A^{\nu}\breve{s}^{2}-\breve{s}^{\nu}(A\cdot\breve{s})\right],
\end{eqnarray}
where the non-linearities (in the electromagnetic quantities) disappear and the equation above corresponds exactly to what we had in the first approach in Eq. (\ref{torsioninducedcurrent}).

%%%%%%%%%%%%%%%%%%%%%%%%%%%%%%%%%%%%%%%%%%%%%%%%%%%%%%%%%%%%%%%%%%%%%%%%
\subsection{Fermions}
%%%%%%%%%%%%%%%%%%%%%%%%%%%%%%%%%%%%%%%%%%%%%%%%%%%%%%%%%%%%%%%%%%%%%%%%
\subsubsection{Fermionic background torsion}

Let us consider first the case in which the matter fields are fermionic spinors. The variation of the Dirac action in a RC space-time (given by the Lagrangian density in Eq.(\ref{eq:mattfer})) with respect to fermionic fields yields the Fock-Ivanenko-Heisenberg-Hehl-Datta equation \cite{Hehl-Data}
\begin{equation}
\label{eq:HehlData}
i\hbar \gamma^{\mu}\tilde{D}_{\mu}\psi-m\psi =
\dfrac{3\kappa^2\hbar^{2}}{8} (\bar{\psi}\gamma^{\nu}\gamma^{5}\psi)\gamma_{\nu}\gamma^{5}\psi \,,
\end{equation}
where torsion was substituted by its source, the spin density of Dirac fermions, via the Cartan equations. Now we introduce electromagnetic fields minimally coupled to torsion, but without backreacting on it. In this case, the variation of the action (\ref{eq:actionEC}) with the new Lagrangian $\mathcal{L}_m=\mathcal{L}_{\Psi}+\mathcal{L}_{\rm Max}$ includes non-minimal couplings of fermions with the four-potential, in the generalized Hehl-Datta equation of EC-Dirac theory. For charged fermions, the new Hehl-Datta equation reads
\begin{eqnarray}
\label{eq:ECDEfer}
i\hbar \gamma^{\mu}\tilde{D}_{\mu}\psi+\left(q\gamma^{\mu}A_{\mu}-\dfrac{\kappa^2\lambda\hbar}{4}f^{\rho}\gamma_{\rho}\gamma^{5}\right)\psi-m\psi\nonumber \\
=
\Bigg(\dfrac{\kappa^{4}\lambda\hbar^{2}}{2}A^{2}+\dfrac{3\kappa^2\hbar^{2}}{4}\Bigg) (\bar{\psi}\gamma^{\nu}\gamma^{5}\psi)\gamma_{\nu}\gamma^{5}\psi 	\nonumber \\
-\dfrac{\kappa^{4}\lambda\hbar^{2}}{2}(\bar{\psi}\gamma^{\beta}\gamma^5\psi)\gamma_{\lambda}\gamma^5\psi A_{\beta}A^{\lambda}.
\end{eqnarray}
The Hehl-Datta term, $\sim\kappa^2\hbar^{2}(\bar{\psi}\gamma^{\nu}\gamma^{5}\psi)\gamma_{\nu}\gamma^{5}\psi$, which is cubic in the spinors, is already present in the usual EC-Dirac theory. This term represents a spin-spin contact interaction inside fermionic matter. For charged anti-fermions, after performing the charge conjugation operation ($\psi\rightarrow -i\gamma^{2}\psi^{*}\equiv \psi^{ch}$) we have instead
\begin{eqnarray}
\label{eq:ECDEafer}
&&i\hbar \gamma^{\mu}\tilde{D}_{\mu}\psi^{ch}-\left(q\gamma^{\mu}A_{\mu}+\dfrac{\kappa^2\lambda\hbar}{4}f^{\rho}\gamma_{\rho}\gamma^{5}\right)\psi^{ch}-m\psi^{ch}
	\nonumber \\
&&\hspace{+0.4cm}=-\Bigg(\dfrac{\kappa^{4}\lambda\hbar^{2}}{2}A^{2}+\dfrac{3\kappa^2\hbar^{2}}{4}\Bigg) (\bar{\psi^{ch}}\gamma^{\nu}\gamma^{5}\psi^{ch})\gamma_{\nu}\gamma^{5}\psi^{ch} 
	\nonumber \\
&&\hspace{+1.75cm}+\dfrac{\kappa^{4}\lambda\hbar^{2}}{2}(\bar{\psi^{ch}}\gamma^{\beta}\gamma^5\psi^{ch})\gamma_{\lambda}\gamma^5\psi^{ch} A_{\beta}A^{\lambda}.
\end{eqnarray}
All cubic terms, similarly to the term having the fermionic charge, have flipped sign after the C-transformation relative to the mass term. This behaviour is connected to the fact that the corresponding effective Lagrangian terms behave in an opposite manner under a C-transformation in relation to the rest of the terms in the Lagrangian \cite{Poplawski:2011xf}.

It has been shown that the Hehl-Datta term, which corresponds to an effective axial-axial spinor interaction of repulsive nature, can provide important physical effects in the particle domain \cite{Khanapurkar:2018jvx,Khanapurkar:2018gyo,Poplawski:2011wj,Poplawski:2011xf,Poplawski:2010jv,Diether:2017oax,Poplawski:2009su}, including a valid mechanism for generating a residual matter/anti-matter asymmetry in the context of baryogenesis in cosmology, and has been shown to posses other applications, such as an effective cosmological constant \cite{Poplawski:2010jv} and non-singular configurations \cite{Poplawski:2009su}.
Such a term can be derived from an effective interaction Lagrangian of the form $L^{\rm int}_{\rm Hehl-Datta}\sim \kappa^{2}\breve{s}^{\mu} \breve{s}_{\mu}$. Analogously, the new cubic terms we have derived also come from similar effective Lagrangian terms quadratic in Dirac's axial (spin) vector $L_{\rm eff}\sim \kappa^{4}\lambda \breve{s}^{2}A^2$, and are induced from the coupling between torsion and the electromagnetic potential. These terms correspond to the quadratic ones appearing in Eq.(\ref{newMaxLagra}). Therefore, the axial-axial or spin-spin contact interaction effect is potentially enhanced (at very high densities) by the presence of the electromagnetic four-potential. Moreover, in general the four-potential propagates, therefore a richer dynamics is induced in the effective spin-spin interaction. This scenario is of course compatible with the fact that we have broken the local (gauge) $U(1)$ invariance under a phase transition above a certain critical value of the spin density. Accordingly, the vector potential that appears explicitly in the dynamical equations can be thought as representing physical degrees of freedom\footnote{In some sense, there are good empirical motivations to consider the electromagnetic potential to represent physical degrees of freedom which come from the interpretations given to the Aharonov-Bohm  effect, namely the observed change in the phase of an electron wave function in the presence of negligible electromagnetic fields, due to the interaction between the fermion and the electromagnetic four-potential \cite{ABohm}.}.

\subsubsection{Full approach, including the contribution of bosonic fields to the spin tensor}

Previously, using Eq. (\ref{newCartan}) we arrived at the fermionic Lagrangian given by
\begin{eqnarray}
\mathcal{L}_{\rm D}&=&\tilde{\mathcal{L}}_{\rm D}-\breve{s}_{\lambda}\breve{s}^{\lambda}\left(\dfrac{3\kappa^{2}}{2}+\lambda\kappa^{4}A^{2} \right)+\lambda\kappa^{4}(A\cdot \breve{s})^{2} 
\nonumber \\
&&+\dfrac{\lambda\kappa^{2}}{2}\epsilon^{\mu\beta\gamma\lambda}\breve{s}_{\lambda}A_{[\mu}\tilde{F}_{\beta\gamma]} \ . 
\end{eqnarray}
If we consider the total matter Lagrangian
\begin{equation}
L_{m}=L_{D}+L_{M}+j^{\mu}A_{\mu},
\end{equation}
including the contribution from the bosonic (electromagnetic) side, we obtain the following extended Dirac (cubic) equation
\begin{eqnarray}
\label{newHehlData}
i\hbar \gamma^{\mu}\tilde{D}_{\mu}\psi&+&\left(q\gamma^{\mu}A_{\mu}-m\right)\psi =f(A)(\bar{\psi}\gamma^{\nu}\gamma^{5}\psi)\gamma_{\nu}\gamma^{5}\psi\nonumber \\
&+&\alpha^{\beta\lambda}(\bar{\psi}\gamma_{\beta}\gamma^5\psi)\gamma_{\lambda}\gamma^5\psi \nonumber \\
&+&\beta^{\alpha}(A,\tilde{F})\gamma_{\alpha}\gamma^5\psi,
\end{eqnarray}
where
\begin{eqnarray}
f(A) &\equiv & \dfrac{3\kappa^{2}\hbar^{2}}{4}+\dfrac{\lambda3\kappa^{4}\hbar^{2}}{2}A^{2}\nonumber \\
\alpha^{\sigma\varepsilon} &\equiv & -\lambda\hbar\kappa^{2}\Big(\dfrac{\kappa^{2}}{2}A^{\sigma}A^{\varepsilon}+\dfrac{1}{2}\Theta^{\lambda\;\;\;\;\varepsilon}_{\;\;\mu\nu}(\epsilon^{\gamma\mu\nu\sigma}
	\nonumber \\
&&+2\lambda\kappa^{2}\epsilon^{\gamma[\mu\vert\rho\sigma}A^{\vert\nu]}A_{\rho}
)A_{\gamma}A_{\lambda}\Big) 
	\nonumber \\
\beta^{\alpha}&\equiv &-\lambda\Big(A_{\lambda}(2A_{\gamma}\Theta^{(\lambda\vert \;\;\alpha}_{\;\;\;\;\mu\nu}T_{M}^{\gamma)\mu\nu}+\tilde{F}_{\mu\nu}\Theta^{\lambda\mu\nu\alpha} ) 
	\nonumber \\
&&+ \dfrac{\kappa^{2}\hbar}{2}\epsilon^{\mu\beta\gamma\alpha}A_{[\mu}\tilde{F}_{\beta\gamma]}\Big) \,,
\end{eqnarray}
and we have
\begin{equation}
\Theta^{\lambda\mu\nu\alpha}\equiv \dfrac{\kappa^{2}\hbar}{4}\left(\epsilon^{\lambda\mu\nu\alpha}+2\lambda\kappa^{2}\epsilon^{\lambda[\mu\vert\rho\alpha}A^{\nu]}A_{\rho} \right), 
\end{equation}
while $T_{M}^{\gamma\mu\nu}$ is the purely bosonic part of the torsion tensor. This equation can be considered in the approximation of space-time flatness and also in the non-relativistic limit. One can then solve the energy levels problem which is expected to reveal a kind of hyperfine structure that could be used to probe for the existence of torsion with high resolution spectrography. In fact, the correction terms in (\ref{newDiracLag}) can be interpreted as effective interaction potentials
\begin{equation}
\mathcal{L}_{\rm D}=\tilde{\mathcal{L}}_{\rm D}+U(\varphi,\chi,\zeta) \ ,
\end{equation}
 with $\varphi\equiv \breve{s}^{2}$, $\chi\equiv A^{2}$, $\zeta\equiv A\cdot\breve{s}$ and we neglected the term linear in $\breve{s}$, for simplicity. Such analysis is currently under study and will be developed in a future work. To close this section, let us mention that for anti-particles we have:
\begin{eqnarray}
\label{newHehlDataanti}
i\hbar \gamma^{\mu}\tilde{D}_{\mu}\psi^{ch}-\left(q\gamma^{\mu}A_{\mu}+m\right)\psi^{ch}=\nonumber \\
-f(A)(\bar{\psi^{ch}}\gamma^{\nu}\gamma^{5}\psi^{ch})\gamma_{\nu}\gamma^{5}\psi^{ch}\nonumber \\
-\alpha^{\beta\lambda}(\bar{\psi^{ch}}\gamma_{\beta}\gamma^5\psi^{ch})\gamma_{\lambda}\gamma^5\psi^{ch} \nonumber \\
+\beta^{\alpha}(A,\tilde{F})\gamma_{\alpha}\gamma^5\psi^{ch}.
\end{eqnarray}
which is not exactly the same dynamics, suggesting possible applications for asymmetries and baryogenesis.

%%%%%%%%%%%%%%%%%%%%%%%%%%%%%%%%%%%%%%%%%%%%%%%%%%%%%%%%%%%%%%%%%%%%%%%%

%%%%%%%%%%%%%%%%%%%%%%%%%%%%%%%%%%%%%%%%%%%%%%%%%%%%%%%%%%%%%%%%%%%%%%%%
\section{Conclusion and discussion} \label{secIV}
%%%%%%%%%%%%%%%%%%%%%%%%%%%%%%%%%%%%%%%%%%%%%%%%%%%%%%%%%%%%%%%%%%%%%%%%

In this work we have studied the Einstein-Cartan-Dirac-Maxwell model with $U(1)$ symmetry breaking and discussed its physical relevance. We considered a Dirac field and an electromagnetic field minimally coupled to torsion, which induces rich gravitational dynamics and non-linear fermionic and bosonic dynamical equations, including non-minimal and self-interactions. We considered two regimes: i) one in which torsion is sourced by fermions and ii) the full case with the contribution from both fermions and bosons to the total spin tensor entering in Cartan's equations.

In general, the effects for the space-time metric only become important at very high (spin) densities, as in the usual EC theory. For example, in the first approach with torsion generated by fermionic spin, torsion (or spin) contributions to the metric field equations scale with $\kappa^{4}\breve{s}^{2}$ for the pure EC correction, while the model with the $U(1)$ symmetry breaking studied here introduces terms both linear and quadratic with torsion, that scale as $\kappa^{4}\lambda \breve{s}\tilde{F}A$ and $\kappa^{6}\lambda \breve{s}^{2}A^{2}$, respectively. This has to be compared with the $\kappa^{2}\tilde{T}_{\mu\nu}$ contribution from the usual stress-energy tensor in GR. Thus, for very strong electromagnetic fields/potential, the term linear in the spin density could become important (in polarized matter) at densities slightly (but not significantly) below Cartan's typical density. On the other hand, the effects of torsion in the electromagnetic and fermionic sectors require a more careful analysis.

Let us discuss the electromagnetic dynamics. The generalized Maxwell theory include terms linear in torsion (also in the spin density) that become significant at densities much lower than Cartan's density, which should be taken into account in strong gravity regimes such as in the interior of astrophysical compact objects (neutron stars, magnetars, quark stars) and in the early Universe. These terms are non-negligible for polarized matter, i.e., for non-random spin distributions and, consequently, the presence of strong magnetic fields  provide the adequate physical conditions for the study of the phenomenology associated with these corrections. For approximately random spin distributions, i.e., for unpolarized matter, only the quadratic terms (in torsion or in the spin density) are non-vanishing with its phenomenology being related to much higher densities. In any case, the presence of strong electromagnetic fields (potential) tend to enhance such effects.

When the $U(1)$ symmetry is broken the corresponding (Noether) charge current is not conserved.  Although the fermionic charge density and number density of the fermions is not conserved locally in this model, the equations suggest interpreting the terms of geometric origin as effective charge currents that compensate and balance the non-conservation of the usual charge current. In other words, by following this interpretation the space-time geometrodynamics would gain physical features, such as effective mass, spin or charge currents, when it couples to matter. When these terms are considered, then  a new conserved quantity is clear. Another way to see this is to deduce the phenomenology associated to such an interpretation and search for possible observational tests of the predictions. In this context, this type of models where the stress-energy tensor or the charge current is not conserved in the usual sense, predict the creation of particles from the energy available in the space-time geometrodynamics, in strong gravity environments.

When the contribution from the bosonic sector to the spin tensor is taken into account, then the bosonic field propagates on a RC spacetime and backreact on its geometry. Since torsion in EC theory is given by an algebraic expression of the matter fields, one then gets non-minimal couplings between these but also self-interactions. Therefore, we obtain effectively a non-linear dynamical equations for the bosonic fields. In fact, just as in the case of fermions where a linear Dirac field in RC space-time of the EC theory is equivalent to a non-linear spinor in GR, also here the linear electromagnetic Lagrangian in the RC space-time leads to an effective non-linear electrodynamics in GR. Non-linear dynamics in the matter fields can emerge naturally from the (minimal) couplings of these fields with the extended space-time geometries of gauge theories of gravity.

In the case of fermionic fields in EC theory, torsion effects can also become significant in environments where the density is lower than Cartan's density. This is not so commonly mentioned in the literature, on the contrary, much emphasis is put on the fact that in EC theory the effects of torsion in Einstein's equations, i.e., for the metric, are only significant at extremely high densities such as those found in the very early Universe or inside black holes. Since the Cartan equations imply $K\sim \kappa^{2}\breve{s}$, after its substitution in the Dirac equation $i\hbar \gamma^{\lambda} D_{\lambda}\psi-m\psi=0$, one obtains the (cubic) Hehl-Datta equation where the torsion-induced term will become significant at (spin) densities comparable to any strong-gravity regime where GR effects become important.

Let us stress that the Hehl-Datta term, which is related to an effective axial-axial (spin-spin) repulsive interaction, has been studied in connection to different physical mechanisms important for particle physics and cosmology, such as non-singular black holes, matter/anti-matter asymmetry and energy-levels, etc. Analogously, in our $U(1)$ symmetry-breaking model similar cubic terms are present that are quadratic in the electromagnetic four-potential. In this case, these torsion-induced corrections scale with $\kappa^{4}$, which means that the corresponding physical effects (on the energy levels, generalized effective Feynmann diagrams, etc) will only become relevant at extremely high densities (Cartan's density or above, but still lower than Planck density). In this model, the minimal coupling  between the electromagnetic potential and torsion induce, at the dynamical equation level, a non-minimal coupling between fermions and electromagnetic potential/fields, in the generalized Dirac equation. Formally, this follows after the substitution of torsion by its corresponding spin density source via Cartan's equations. The new terms are both linear and cubic in the spinors. The former introduces effects that will become relevant around the same densities as for the original Hehl-Datta term. These considerations motivate further study on the full EC-Dirac-Maxwell dynamics inside astrophysical compact objects.

Finally, let us mention several cosmological, astrophysical and particle physics applications that can be worked out from the theory considered in this work. In Cosmology one expects the possibility of non-singular models as in the usual EC model, and new physics during the torsion-dominated era. One should also expect the production of gravitational waves from the transitions between primordial phases: from the $U(1)$-broken phase to the $U(1)$-restored phase, and from the usual torsion-dominated phase of EC to the radiation phase. These transitions can contribute to a stochastic gravitational wave background of cosmological origin, with possible imprints from the physics beyond the standard model. 

On the other hand, the standard EC theory can prevent black hole singularities and, therefore, the research on whether one can have equilibrium configurations in compact objects denser than neutron stars, before the appearance of an horizon, is of utmost relevance. In our model we have physical mechanisms induced by torsion that act as an effective repulsive interaction, which could possibly provide the required pressure to balance the self-gravity of a newly born (unstable) neutron star. After the coalescence of two neutron stars in models of GW emission, it is usually assumed that the resulting object stabilizes to a neutron star or decays into a black hole (directly or after some relaxation time), due to GR instabilities, but in modified gravity, torsion/spin effects should allow for other equilibrium configurations, i.e, stable compact objects denser than neutron stars.    

To conclude, it is necessary to investigate whether the astrophysical data about the final object might be reinterpreted using models with torsion. In our view, there are good motivations to consider gravitational models where non-Riemannian geometries, fermionic spin densities, and phase transitions become important, which can be tested with astrophysical, cosmological and gravitational wave observations.  Work along these lines is currently underway.

%%%%%%%%%%%%%%%%%%%%%%%%%%%%%%%%%%%%%%%%%%%%%%%%%%%%%%%%%%%%%%%%%%%%%%%%
\section*{Acknowledgements}
%%%%%%%%%%%%%%%%%%%%%%%%%%%%%%%%%%%%%%%%%%%%%%%%%%%%%%%%%%%%%%%%%%%%%%%%

FC is funded by the Funda\c{c}\~ao para a Ci\^encia e a Tecnologia (FCT, Portugal) predoctoral grant No.PD/BD/128017/2016.
DRG is funded by the \emph{Atracci\'on de Talento Investigador} programme of the Comunidad de Madrid, No.2018-T1/TIC-10431, and acknowledges support from the FCT research grant PTDC/FIS-PAR/31938/2017,  the projects FIS2014-57387-C3-1-P and FIS2017-84440-C2-1-P (MINECO/FEDER, EU), the project SEJI/2017/042 (Generalitat Valenciana), the Edital 006/2018 PRONEX (FAPESQ-PB/CNPQ, Brazil) and the
EU COST Actions CA15117 and CA18108. 
FSNL acknowledges support from the Scientific Employment Stimulus contract with reference CEECIND/04057/2017. 
The authors also acknowledge funding from FCT projects No.UID/FIS/04434/2019 and No.PTDC/FIS-OUT/29048/2017.
Finally, the authors thank the anonymous referee for extremely helpful comments and suggestions.

%%%%%%%%%%%%%%%%%%%%%%%%%%%%%%%%%%%%%%%%%%%%%%%%%%%%%%%%%%%%%%%%%%%%%%%%

%%%%%%%%%%%%%%%%%%%%%%%%%%%%%%%%%%%%%%%%%%%%%%%%%%%%%%%%%%%%%%%%%%%%%%%%
\end{document}